\documentclass[conference]{IEEEtran}
\usepackage{cite}
\usepackage{amsmath,amssymb,amsfonts,multirow,array}
\usepackage{algorithmic}
\usepackage{textcomp}
\usepackage{tabularx}
\usepackage{xcolor}
\usepackage{bm}
\usepackage{graphicx}
\ifCLASSOPTIONcompsoc
    \usepackage[caption=false, font=normalsize, labelfont=sf, textfont=sf]{subfig}
\else
\usepackage[caption=false, font=footnotesize]{subfig}
\fi\usepackage{url}
\usepackage[T1]{fontenc}
\def\BibTeX{{\rm B\kern-.05em{\sc i\kern-.025em b}\kern-.08em
    T\kern-.1667em\lower.7ex\hbox{E}\kern-.125emX}}
\begin{document}

\title{BeamLearning: an end-to-end Deep Learning approach for the angular localization of sound sources using raw multichannel acoustic pressure data}

\author{\IEEEauthorblockN{Hadrien Pujol}
\textit{LMSSC, Cnam,}\\
Paris, France \\
hadrien.pujol@lecnam.net
\and
\IEEEauthorblockN{\'Eric Bavu }
\textit{LMSSC, Cnam,}\\
Paris, France \\
eric.bavu@lecnam.net
\and
\IEEEauthorblockN{Alexandre Garcia }
\textit{LMSSC, Cnam,}\\
Paris, France \\
alexandre.garcia@lecnam.net}

\date{\today}

\maketitle

\begin{abstract}

\phantom{bla}\\

\textit{The following article has been submitted to the special issue on Machine Learning in Acoustics in JASA. After it is published, it will be found at \url{http://asa.scitation.org/journal/jas}. }\\

Sound sources localization using multichannel signal processing has been a subject of active research for decades. In recent years, the use of deep learning in audio signal processing has allowed to drastically improve performances for machine hearing. This has motivated the scientific community to also develop machine learning strategies for source localization applications. In this paper, we present BeamLearning, a multi-resolution deep learning approach that allows to encode relevant information contained in unprocessed time domain acoustic signals captured by microphone arrays. The use of raw data aims at avoiding simplifying hypothesis that most traditional model-based localization methods rely on. Benefits of its use are shown for realtime sound source 2D-localization tasks in reverberating and noisy environments. Since supervised machine learning approaches require large-sized, physically realistic, precisely labelled datasets, we also developed a fast GPU-based computation of room impulse responses using fractional delays for image source models. A thorough analysis of the network representation and extensive performance tests are carried out using the BeamLearning network with synthetic and experimental datasets. Obtained results demonstrate that the BeamLearning approach significantly outperforms the wideband MUSIC and SRP-PHAT methods in terms of localization accuracy and computational efficiency in presence of heavy measurement noise and reverberation.\\
\end{abstract}


\section{\label{sec:1} Introduction}

Sound sources localization (SSL) is a research field in acoustics that has only recently made use of machine learning approaches. Even if research works on the subject can be found as early as in the 1990s\cite{steinberg1991neural} -- and sporadically until 2015\cite{Guentchev1998learning, weng2001three, talmon2011supervised, deleforge2015acoustic} -- it is mainly from 2017 onwards, in particular with the development of deep learning, that the scientific community proposed to use advanced machine learning techniques for SSL tasks. The present paper proposes a supervised deep learning approach for the localization of acoustic sources in aerial environment, using raw, unprocessed time-domain acoustic pressure measurements on a compact array of microphones. Although not aiming at giving a systematic and exhaustive bibliographical review on this subject, this section presents an overview of the methods recently proposed in the scientific literature on this subject in which motivates the development of the BeamLearning approach.

The growing interest in using deep learning techniques for SSL has been recently illustrated by the publication of the work of several authors\cite{chakrabarty2019multi, perotin2019crnn, adavanne2018sound, brendel2019distributed} in a special issue on acoustic source localization\cite{gannot2019introduction}. The LOCATA challenge\cite{lollmann2018locata}, which brings together a corpus of data recorded by different microphone arrays for different scenarios (single or multiple, moving or static sound sources combinations recorded on static or moving arrays), has also motivated the proposal of the use of a deep neural network \cite{pak2018locata}, which has been shown to outperform both the baseline method and the conventional MUSIC method. While most publications on the subject use supervised learning strategies\cite{liu2020source, pak2019sound, comanducci2020source, yasuda2020sound, varzandeh2020exploiting, chakrabarty2019multi, perotin2019crnn, brendel2019distributed,gemba2019robust, liu2020multi, chakrabarty2017broadband, adavanne2018sound, Adavanne2019_DCASE, hirvonen2015classification,he2018deep,sundar2020raw,yalta2017sound,ma2017exploiting,nguyen2018autonomous,sivasankaran2018keyword,vesperini2016neural,ferguson2018sound, tang2019regression,salvati2018exploiting,takeda2016sound,yiwere2017distance, xiao2015learning,suvorov2018deep, vera2018towards, huang2020time,comminiello2019quaternion} for audible acoustic sources in air and for underwater localization\cite{gemba2019robust, liu2020multi}, unsupervised or semi-supervised learning approaches have also recently been successfully applied for SSL tasks in reverberant and noisy environments \cite{hu2020semi, hu2020unsupervised}. Among these studies, the majority of authors apply the SSL task to a single active source -- this is also the case in this paper -- while others evaluate their methods for multiple sources localization\cite{chakrabarty2017broadband, adavanne2018sound, Adavanne2019_DCASE, hirvonen2015classification,he2018deep, sundar2020raw} or propose specific deep learning approaches to count the number of active sources\cite{stoter2018countnet,GrumiauxKGG20}.\\

The rise of deep learning algorithms based on convolutional neural networks for machine hearing and acoustics\cite{bianco2019machine} -- along with their ability to improve the robustness of direction of arrival (DoA) estimation techniques in presence of noise and reverberation -- led to the use of  various kinds of inputs to the proposed neural network architectures. Among them, the most common choices are based on time-frequency representations based on short-time Fourier transforms\cite{chakrabarty2017broadband,chakrabarty2019multi,adavanne2018sound,he2018deep}. Using this bidimensional representation, some authors proposed to exploit the phase information only\cite{chakrabarty2017broadband}, others used both amplitude and phase\cite{adavanne2018sound, Adavanne2019_DCASE}, and some used the power computed from the STFT matrix\cite{yalta2017sound, hirvonen2015classification, salvati2018exploiting}. Still, other kinds of imput features have been proposed in the litterature, such as the GCC features\cite{xiao2015learning}, the eigenvectors of the spatial covariance matrix\cite{takeda2016sound, yiwere2017distance}, or ambisonics signals\cite{adavanne2018sound,comminiello2019quaternion, perotin2019crnn, tang2019regression}. However, there is still no consensus on the best representation to use in order to better encode the information needed to localize sound sources.\\

Since the unprocessed, time-domain audio signals contain all the information to be extracted, the scientific community has recently put some efforts to directly use the raw waveforms as inputs for deep learning models, either for machine hearing tasks\cite{dieleman2014end,dai2017very,sainath2015learning,lee2018samplecnn,sainath2015learning,bavu2019timescalenet,ravanelli2018speaker} or for SSL tasks\cite{suvorov2018deep, vera2018towards, huang2020time, sundar2020raw}. Joint acoustic model learning from the raw waveform has therefore emerged as an active area of research in the last few years, and recent works have shown that this approach, also known as \textit{end-to-end learning}, allows to successfully learn the temporal and spatial dynamics scales of the waveforms. These studies, along with recent advances in machine learning architectures for one-dimensional signals\cite{oord2016wavenet,kaiser2017depthwise,rethage2018wavenet} and the work of the authors on a similar approach for speech and environmental sounds recognition\cite{bavu2019timescalenet} has motivated the present work, which aims at showing the benefit of a deep-learning multi-resolution approach for the SSL task, that allows to avoid the need to pre-process the multichannel waveforms in order to encode the relevant information contained in raw acoustic measurements.\\

Among the already published studies on SSL using deep learning techniques, the problem is the most commonly treated in a classification framework, where the space is divided into distinct zones. These zones can be angular sectors, and thus represent portions of azimuthal angles\cite{yalta2017sound, ma2017exploiting, nguyen2018autonomous, sivasankaran2018keyword}, or combine azimuthal and elevation angles to define portions of a sphere\cite{Adavanne2019_DCASE}. Perotin et al.\cite{perotin2019crnn} proposed to use a classification task to also estimate the distance between the source and the microphone array. However, for a SSL task, a regression approach seems as legitimate as classification\cite{perotin2019regression}. When a high angular precision is required, a classification approach may also not be longer sufficient, and a regression approach, giving a continuous numerical value of the source's angular position\cite{vesperini2016neural,ferguson2018sound,tang2019regression} even appears to be more accurate in scenarios with diffuse interference\cite{perotin2019regression}. In a recent study\cite{sundar2020raw}, the authors propose to combine the two approaches in order to determine the position of several speakers in a room. The position is first roughly inferred using a classification approach, and then refined using a regression approach. With regard to the BeamLearning network detailed in the present paper, both approaches will be studied in order to compare the observed results.\\

Machine learning approaches for SSL localization require a large amount of training data. Since signals captured by microphone arrays with different geometries are radically different, individual experimental data collection is required for each specific type of microphone array. This is why most SSL studies use simulated data, which therefore need to be as realistic as possible in order to adapt to real recording conditions. Some solutions have been proposed in the litterature using domain adaptation\cite{he2019adaptation} when the mismatch between training and testing conditions is too high. We therefore also developed an efficient tensor GPU-based computation of synthetic room impulse responses using fractional delays for image source models.\\

In this article, we present the Beamlearning approach, which relies on the use of a deep neural network (DNN) DoA estimation system, using raw multichannel measurements. The BeamLearning neural network architecture has been specifically developed in order to be insensitive to the microphone array geometry and to the number of sensors involved in the measurement array, and to be compatible with both classification and regression approaches. The obtained results are evaluated for both experimental and synthetic data using speech signals, in anechoic and reverberant environments, with strong noisy conditions. These results are compared on the same data with two state of the art model-based SSL methods (MUSIC and SRP-PHAT). Section II provides details on the method used to build the synthetic datasets, and the BeamLearning neural network architecture is extensively detailed. The different datasets used in this paper are described in Section III, along with the evaluation methods and the training procedure. The ensemble of three multichannel, multiresolution learnable stacked filterbanks, which constitutes the core of the first subnet of the BeamLearning network is analyzed in detail in Section IV. Finally, several SSL experiments are presented and analyzed in Section V, before drawing our conclusions in Section VI.

\section{\label{Methods}Methods}

\subsection{\label{dataset_generation}Synthetic and experimental datasets generation}

For both the dataset generation and the neural network architecture, the BeamLearning approach proposed in the present paper has been tailored and extensively tested in order to accomodate to any microphone array configurations in arbitrary reverberant rooms. However, in this paper, for sake of clarity and concision, we chose to illustrate the results using only a particular microphone array. For the experimental dataset, we used the MiniDSP\textsuperscript{\small{\textregistered}} UMA-8 compact microphone array (\url{https://www.minidsp.com/products/usb-audio-interface/uma-8-microphone-array}). This compact array is composed of $N_c = 7$ digital MEMS microphones. The first one is placed at the center of the circular array, and the 6 others microphones  are evenly distributed over the perimeter of a 8 cm diameter circle. This geometry is mostly similar to the microphone arrays used in personal assistants such as Amazon Echo\textsuperscript{\small{\textregistered}}, or Apple HomePod\textsuperscript{\small{\textregistered}}, where the estimation of a speaker's direction could represent an important pre-requisite in the processing chain for hands-free communication during voice interaction with these devices in noisy and reverberating environments.\\

The experimental datasets have been generated using the UMA-8 microphone array, placed at the ''sweet spot'' of the laboratory's 5-th order ambisonic playback system\cite{lecomte2016fifty}, which allows a great flexibility and reproducibility for the generation and labelling task of experimental datasets. This sound field synthesis system consists in a 1.07 m radius spherical loudspeaker array following a 50-node Lebedev grid. The loudspeakers are made with a tubular plastic cabinets and 2''~drivers (AuraSound\textsuperscript{\small{\textregistered}} NSW2-380-8A). Datasets of 72000 sources positions (80\% for learning, 10\% evaluation and 10\% for testing) have been encoded in the ambisonic domain in order to drive the 50 loudspeakers\cite{lecomteambitools}. These sources positions are randomly generated using the procedure detailed in \ref{subsec:Datasets}.\\

In this paper, the BeamLearning network has also been trained and tested using synthetic datasets corresponding to different room configurations. For these datasets, the microphone array geometry matches the one used in the experimental datasets. Since there is a need to model accurately a large number of sound sources positions and signals in various simulated environments, we specifically developed for this purpose a GPU-based computation of synthetic room impulse responses (RIRs) using fractional delays\cite{pujol2019source}. This computation of realistic RIRs relies on the use of the image source method\cite{allen1979image}. In this process, all the reflections on the room's boundaries are computed, for any time of arrival on the array smaller than the reverberation time (RT) of the room. As an example, for a typical classroom size of $7{ }\times10{}\times3.70$~m with a RT of 0.5~s, the whole number of image sources requires a high order of reflections (up to the 60th order), and represents more than 80000 image sources for each of RIR of the synthetic dataset. For each simulated environment, 48000 primary source positions are used in the RIR computations. These sources positions are randomly generated using the procedure detailed in \ref{subsec:Datasets}. Using the UMA-8 microphone array, the dataset therefore corresponds to the computation of 336000 RIRs for each room. For each of these RIRs, the whole number of image sources positions and corresponding attenuations that contribute to the acoustic field for a time interval of $[0;T_R]$ are computed using the Pyroomacoustics library\cite{pyroomacoustics}. For such a large number of RIRs and image sources, the existing frameworks did not allow to perform the RIR computation efficiently and accurately enough. This is the reason why we developed for this specific application a fast parallel batch RIR computation performed on the GPU using the Tensorflow APIs\cite{tensorflow2015-whitepaper}. This implementation is achieved using sparse tensors computations and an efficient fractional delay filtering implementation\cite{pujol2019source}. For a compact microphone such as the UMA-8, there is indeed a critical need to keep the precision of the time delays corresponding to the distance between each of the image sources involved in the RIR computation and the microphones. In order to implement the RIR with non integer sample delays, we therefore used a 7\textsuperscript{th} order Lagrange interpolation\cite{smithdigitalfilters}, that allows to perform a better accuracy than standard truncated sinc interpolations, with a much lower computation cost. The integer part of the sample delay is represented as a sparse array, and the 8 coefficients of each fractional delay filters along with the individual dampings corresponding to the cumulative effect of each wall of the room and the distance between the image source and the microphones are then used to compute the RIRs using every source image contributions\cite{pujol2019source}. Using this large number of RIRs, the complete dataset is built using batch convolutions with real-life recordings for each simulated environments.

\subsection{\label{subsec:beamlearning_architecture}Proposed neural network architecture}

\begin{figure*}[ht!]
		\centering
\includegraphics[width=0.99\textwidth]{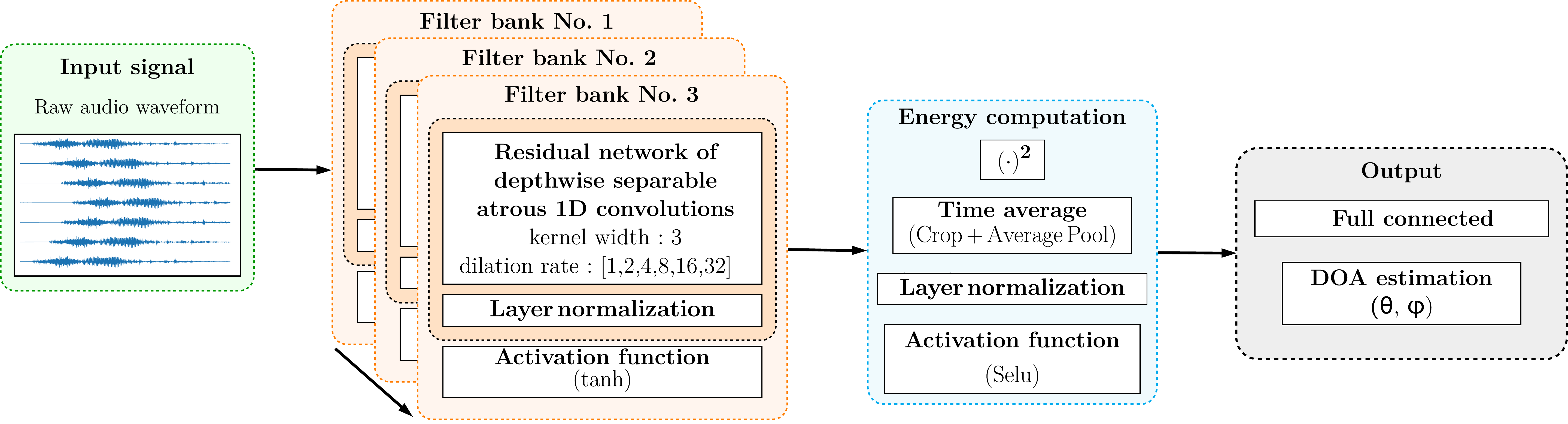}
	\caption{\label{fig:Reseau}{Schematical representation of the global architecture of the BeamLearning network. This neural network takes a raw multichannel waveform measured by a microphone array, and outputs the DOA estimation of the emitting source.}}
	
\end{figure*}

The proposed BeamLearning neural network is composed of three subnetworks, which are further detailed in the following subsections. This DNN architecture is fed with raw multichannel audio data measured by a microphone array (see Fig.~\ref{fig:Reseau}), without any further preprocessing. This allows to perform a joint feature learning along with the SSL task, and to avoid manual input features choices that may not be well adapted to particular microphone array geometries. We also specifically developed this architecture in order to allow real-time SSL inference after convergence of the training process. This led to the use of input frames of length $N_t = 1024$ samples. Since most conventional beamforming methods can be thought as filter and sum approaches\cite{brandstein2013microphone}, the global neural architecture depicted on Fig.~\ref{fig:Reseau} has been developed in a similar way. The DNN architecture is mainly based on the use of a learnable filterbank, with specific operations allowing the neural network to be expressive enough to achieve the SSL task in reverberant and noisy environments using raw multichannel audio input data of dimension $N_c \times N_t$, with $N_c = 7$ for the particular array used in this paper.\\

 The first subnetwork can be thought as a set of finite impulse-response (FIR) filters with learnable coefficients, that self-adapts to the multichannel audio input data for the SSL task. The convolutional neural cells, which are widely used in DNN architectures, can indeed be thought as a strict equivalent to learnable finite impulse response (FIR) filters in standard digital signal processing (DSP). For this purpose, we propose the use of residual networks of one-dimensional depthwise separable atrous convolutions, which allow to operate in a computationally efficient way, with a similar strategy to the one used in the FrameNet subnet that we proposed in a previous published study for sound recognition\cite{bavu2019timescalenet}. A depthwise separable convolution consists in a convolution performed independently over each channel of an input, followed by a pointwise convolution, \emph{i.e.} a 1x1 convolution, projecting the channels output by the depthwise convolution onto a new channel space. This process allows to operate a convolution on data, faster, with much less parameters than standard convolutions\cite{chollet2017xception}. This residual subnetwork of depthwise separable atrous convolutions is described in detail in \ref{subsec:filterbank_subnetwork}. The overall output of these three filterbanks -- which are composed of $3\times768$ cascaded learnable filters, each of them having a channel multiplier of 4 -- is a two-dimensional map, where the first dimension is $N_t$, and the second dimension represents the $N_f$ outputs of the cascaded learnable filters. In the present paper, for a 2D-DoA finding task, $N_f = 128$. This hyperparameter value has been determined after extensive preliminary training and testing phases for the proposed DOA task, in order to optimize the neural network performances while limiting the impact on the computational cost on a single GPU. The obtained data representation is then fed to the second subnetwork depicted on Fig.~\ref{fig:Reseau}, which aims at computing an energy-like representation in $N_f$ dimensions. This subnet is inspired by Beamforming approaches, where the sound source position is determined using a maximization of the Steered Response Power (SRP) of a beam former\cite{brandstein2013microphone}. The last part of the BeamLearning network aggregates the energy of the $N_f$ channels using a full connected subnetwork in order to infer the sound source position. In the following, the architecture of each of these sub-networks is presented and discussed in relation to the physics of the associated SSL problem.\\

 \subsubsection{\label{subsec:filterbank_subnetwork}Learnable filterbanks : raw waveform processing}\
 
\begin{figure}
	\centering
	\includegraphics[width=0.9\columnwidth]{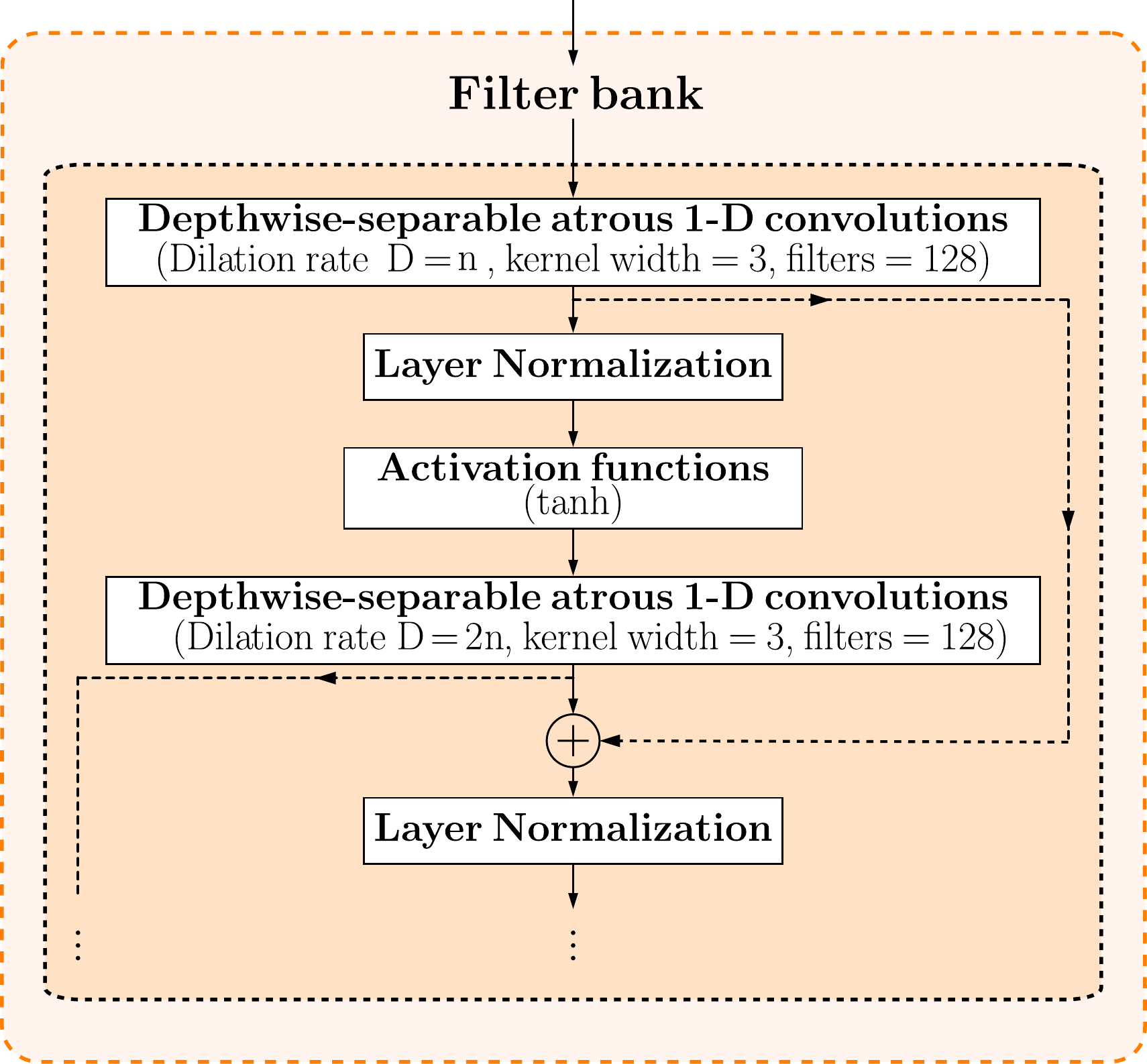}
	\caption{\label{fig:Filtre}{Inner architecture of a learnable filterbank module, showing the residual connections (dotted lines) between the successive layers. Each layer corresponds to a different dilation rate D, taking values 1,2,4,8,16, and 32. In each depthwise-separable convolutional layer, the depthwise convolution is achieved using a channel multiplier $m=4$, and a pointwise convolution mapping the outputs of the depthwise convolution into $N_f = 128$ pooled channels.}}
	
\end{figure}

As introduced in the previous subsection, from a machine-learning point of view, the first subnetwork in the BeamLearning DNN consists of three stacked modules, which are residual networks of one-dimensional depthwise separable atrous convolutions. The proposed filterbank module architecture is detailed on Fig.~\ref{fig:Filtre}. The use of convolutional layers is a common operation in deep learning for audio applications, that allows to extract features from the data. From a signal processing point of view, the kernel of the convolutional cells can be seen as the Finite Impulse Response (FIR) of a learnable filter. However, acoustic model learning from the raw waveform using dense kernels can lead to large FIR sizes in order to be efficient at filtering at low frequency\cite{bavu2019timescalenet,ravanelli2018speaker}. As a consequence, rather than making convolutions with large and dense kernels, the proposed filterbank is implemented using one-dimensional atrous convolutional kernels, which have a long history in classical signal processing, and have originally been developed for the efficient computation of the undecimated wavelet transform\cite{holschneider1990real}. The use of this kind of convolution with upsampled filters has been revisited in the context of Deep Learning and have already been shown as an efficient architecture for audio generation\cite{oord2016wavenet}, denoising\cite{rethage2018wavenet}, neural machine translation\cite{kaiser2017depthwise}, and word recognition\cite{bavu2019timescalenet}. Similarly to previous published studies\cite{oord2016wavenet,rethage2018wavenet,bavu2019timescalenet}, we use dilation rates which are multiplied by a factor of two for each successive layers and very short kernel widths of 3 samples (early experiments with larger filters of length 5, 7 and 9 showed inferior performances). This allows to achieve a large receptive field (127 samples for a single residual subnetwork of depthwise separable atrous convolutions, and a total receptive field of 379 samples for the three stacked learnable filterbanks), with only $3\times6$ sets of one-dimensional convolutions with kernels of size $1\times3$, corresponding to dilation rates ranging from 1 to 32.\\

 The proposed network topology has been developed in order to optimize the SSL performances in the frequency range of [100~Hz ; 4000~Hz]. For such a large frequency range, using raw audio data, it is useful to offer different timescales of analysis in order to extract the relevant information for the SSL task. The use of three stacked residual atrous convolutional filterbanks depicted on Fig.~\ref{fig:Reseau} and Fig.~\ref{fig:Filtre} -- which share the same architecture but do not operate on the same data, being in cascade -- allow the network to operate on multiple time scales in the range of [68 $\mu$s~;~8.6 ms]. The corresponding filter impulse response lengths can thus take values of one period for frequencies ranging from 115~Hz to 14.7~kHz without impacting the computational efficiency. Since the frequency response of a FIR filter is highly dependent on its impulse response length, the proposed network topology -- along with the skip connections offered by the residual network -- allows for a wide range of filtering possibilities in the targeted frequency range.\\

In our approach, we use non-causal depthwise separable convolutions, which present the considerable advantage of making a much more efficient use of the parameters available for representation learning than standard convolutions\cite{kaiser2017depthwise}. The convolutions are performed independently over channels (depthwise separable convolutions), with a channel multiplier $m$. These computed depthwise convolutions are then projected onto a new channel space of dimension $N_f$ for each layer using a pointwise convolution, which is a type of convolution that uses a kernel of size $1\times1$ : this kernel iterates through every single time sample of the input tensor. This kernel has a depth of however many channels its input tensor has. From a signal processing point of view, this approach aims at pooling together the contents in the filtered multichannel soundwave that share similar spatial and spectral features, in order to ease the sound localization task: the pointwise convolution therefore helps combining the pooled channels in order to enhance the expressivity of the network.\\

As shown on Fig.~\ref{fig:Reseau}, three of these filterbank modules are stacked, and residual connections are added between each layers of the three successive filterbanks (see Fig.~\ref{fig:Filtre}), in order to allow shortcut connections between layers and offer increased representation power by circumventing some of the learning difficulties introduced by deep layers\cite{he2016identity}. These skip connections allow the information to flow across the layers easier by bypassing the activations from one layer to the next. This limits the probability of saturation or deterioration phenomenons of the learning process, both for forward and backward computations in deep neural networks\cite{he2016identity,he2016deep,srivastava2015training}.\\

As depicted on Fig.~\ref{fig:Filtre}, each atrous convolutional layer is followed by a layer-normalization layer (LN)\cite{ba2016layer}, which allows to compute layer-wise statistics and to normalize the nonlinear tanh activation accross all summed inputs within the layer, instead of within the batch with a standard batch-normalization\cite{ioffe2015batch, ioffe2017batch} process. The layer-normalization approach has the great advantage of being insensitive to the mini-batchsize\cite{ba2016layer}, and to be usable even if the batch dimension is reduced to one, when the DNN is used for realtime inference. Each LN layer is then followed in this module by a tanh nonlinear activation function, which was selected after extensive comparison with other standard activation functions for the SSL task using the BeamLearning architecture.\\

\subsubsection{Pseudo-energy computation}\

In the following, the set of $N_f$ time-domain outputs $s^{(i)}[n]$ of the filterbank subnetwork will be denoted as $\mathbf{S}_{i,n}$ - where $i$ stands for the filter channel index, and $n$ for the time sample - the bold notation signifying that this a two-dimensional tensor.  $\mathbf{S}_{i,n}$ is fed to the second subnetwork of the BeamLearning DNN, which allows to compute a pseudo-energy for each of the $N_f$ pooled channels. As shown in Fig.~\ref{fig:Reseau},  from a machine learning point of view, $\mathbf{S}_{i,n}$ is squared, and cropped in the time dimension -- therefore only keeping the time frames corresponding to the valid part for all the atrous convolutional layers used in the filterbanks subnetwork -- and undergoes an averagepool operation. This pooling operation acts as a dimensionality reduction and has been preferred to standard maxpool operation, since the proposed pseudo-energy can be seen as an equivalent to a mean quadratic pressure, which is similar to the value that is maximized by standard model-based Beamforming algorithms\cite{brandstein2013microphone}. The output of this deterministic pseudo-energy computation is then fed to a LN layer\cite{ba2016layer} in order to normalize the Selu\cite{klambauer2017self} nonlinear activations, which in this case helps to speed up the convergence process.\\

\subsubsection{\label{subsec:output}Matching the pseudo-energy features to the source angular output}\

The last layer of the BeamLearning network is a standard full connected layer, which aims at combining the previously computed pseudo energy features in order to infer the source DoA. In our experiments, the BeamLearning approach is evaluated for 2D DoA determination tasks using either a classification framework or a regression framework. For the classification problem experiments, the BeamLearning DNN output is a n-dimensional vector representing the probabilities of the source belonging to $n$ different angular partitions. For a regression problem, and in agreement with Adavanne et al.\cite{adavanne2018sound} proposal, the output corresponds to the projection of the source position on the unit circle (in 2D) or on the unit sphere (in 3D). This output vector therefore corresponds to the unit vector pointing towards the source DOA. Because of the angular periodicity of the DoA determination problem, this allows an easier retro-propagation process  of errors from these projections, than directly from the error of a periodic angle. The angular output DoA $\theta$ (and if needed $ \phi$ in 3D) is then deduced from this projected output.\\

\subsubsection{\label{subsec:trainig} Training procedure}\

Using a classification framework, the BeamLearning DNN is trained with one-hot encoded labels corresponding to the angular area where the source is located, therefore allowing to compute the cross-entropy loss function between estimated labels and ground truth labels. For the regression problem experiments, the BeamLearning DNN is trained with labels corresponding to the projection of the source position on the unit circle, therefore allowing to compute a standard L2-norm loss function between estimated labels and ground truth labels, which represents a geometrical distance between the prediction and the true DOA using cartesian coordinates. \\

The DNN variables updates and the back propagation of errors through the neural network are optimized using the Adaptive Moment Estimation (Adam\cite{kingma2014adam}) algorithm, which performs an exponential moving average of the gradient and the squared gradient, and allows to control the decay rates of these moving averages. In addition to the natural decay of the learning rate that Adam performs during the learning process, we set an exponential learning rate decay. The learning rate $\lambda$ for each step is set to decrease exponentially at each learning iteration $k$, from a maximum value $\lambda_{Max} = 10^{-3}$ to a minimum value $\lambda_{min} = 10^{-6} $ : 
$$\lambda = \lambda_{min} + (\lambda_{Max} - \lambda_{min}) * e^{-\dfrac{k}{N_{iteration}}}$$ The models have been implemented and tested using the Python Tensorflow\cite{tensorflow2015-whitepaper} open source software library, and computations were carried out on a Nvidia\textsuperscript{\small{\textregistered}} GTX 1080Ti GPU card, using mini-batches of 100 raw multichannel waveforms for each training steps. All the network variables were initialized using a centered truncated normal distribution with a standard deviation of 0.1. Using the proposed datasets, the convergence of the training process requires approximately 150 epochs using a Nvidia\textsuperscript{\small{\textregistered}} GTX 1080Ti GPU card.\\

\section{\label{sec:Evaluation}Evaluation}	

\subsection{\label{subsec:Datasets}Datasets}

In the present paper, we evaluate the performances of the proposed BeamLearning DNN for a 2D DoA SSL task, using experimental and synthetic datasets. Both dataset generation methods have been detailed in subsection \ref{dataset_generation}. In the present section, the source positions around the microphone array used in the datasets are detailed, and the different propagating environments are also presented.\\

\begin{figure}[ht]
	\centering
	\includegraphics[width=0.9\columnwidth]{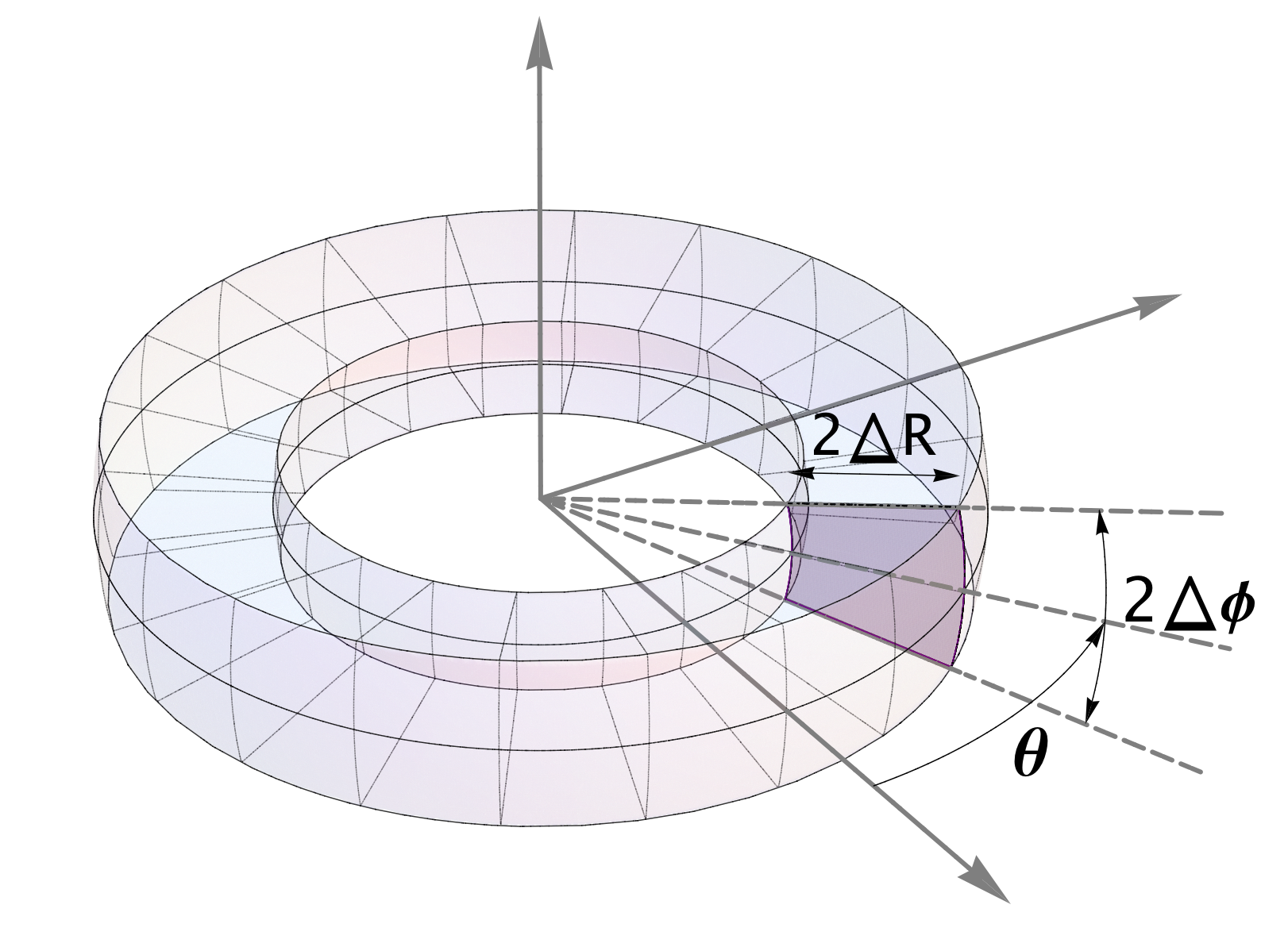}
	\caption{\label{fig:Pos_src}{Sound sources positioning strategy around the microphone array for the generation of the experimental and synthetic datasets.}}
	
\end{figure}

In the following, the source positions are denoted using spherical coordinates (R, $\theta, \phi$) centered on the microphone array, where R is the distance to the central microphone, and ($\theta, \phi$) are the azimuthal and elevation angles defined in Fig.~\ref{fig:Pos_src}. In the present study, since we evaluate the BeamLearning approach for a 2D DoA determination task, the $\theta$ coordinate is the only DNN global output. However, in order to increase the robustness to to spatial variations, the source positions are randomly picked from a uniform distribution in a torus of section $ R \times (2 \Delta R) \times \sin(2 \Delta \phi)$ and of central radius $R$, as depicted in Fig.~\ref{fig:Pos_src}. 

For each synthetic dataset, a set of 48000 sources positions is randomly generated using this procedure. For the experimental dataset, a set of 72000 sources positions has also been randomly generated using the same geometry. The datasets are split into training, validation and test sets in the standard ratios of 80:10:10.\\

  In this following, the BeamLearning DNN is trained and tested using synthetic data in two different reverberating environments, and using experimental data in a real room. The first synthetic dataset  is generated in a simulated room corresponds to a typical classroom size of $7\times10\times3.7$~m and a reverberation time of 0.5~s. In this case, the microphone array is positioned at coordinates $[4,6,1.5]$~m, the origin of the Cartesian coordinate system corresponding to a corner of the room. The second simulated room exhibits an almost cubic shape, with dimensions $5\times5\times4$~m, and a reverberation time $T_r = 0.5$~s. For both simulated rooms, the room boundaries exhibit homogeneous absorbing coefficients computed using the method proposed by Lehmann et al.\cite{lehmann2007reverberation,lehmann2008prediction} in order to ensure a simulated reverberation time of 0.5~s. In order to also study the localization performances offered by the proposed BeamLearning approach in presence of measurement noise independantly of reverberation, a free field dataset has also been generated for this particular experiment.\\

For the synthetic datasets, a dataset of multichannel RIRs is first computed for each propagating environment using the method detailed in \ref{dataset_generation}. These sets of multichannel RIRs are then convoluted with several kinds of signals emitted from the position of the simulated sound sources. This common method\cite{vera2018towards, perotin2019crnn, Adavanne2019_DCASE} allows a great flexibility in terms of emitted signals in a memory-efficient way. In the present study, the signals used to generate the synthetic datasets using the computed RIRs are composed of anechoic recordings of symphonic music\cite{lokki2008recording} (\url{https://users.aalto.fi/~ktlokki/Sinfrec/sinfrec.html}), outside recordings of multiple women talking in Danish (\url{https://odeon.dk/downloads/anechoic-recordings/}), and car horn honks from the UrbanSound8K dataset\cite{Salamon:UrbanSound:ACMMM:14}.\\

During the training phase, each multichannel recordings of the training mini batches are added to a random multichannel Gaussian noise with randomly picked signal-to-noise ratio (SNR) values in the range of $[x,+\infty[$ dB. This process can be interpreted as a data augmentation strategy, which allows to ease the generalization capabilities of the proposed method during the learning process. In the following, the presented results will include data augmentation strategies during the training phase with $x = 5$~dB (section \ref{sec:Results}), $x = 20$~dB (section \ref{sec:Output}), $x = 25$~dB (section \ref{subsec:Robustness_noise}), and $x = +\infty$~dB (without any data augmentation, section \ref{subsec:Robustness_noise}). The proposed data augmentation strategy will be shown to offer an increase of robustness to measurement noise in comparison to state of the art SSL algorithms.\\

 In order to evaluate the 2D-DoA SSL performances of the BeamLearning approach and to compare it in a reproducible way to model-based algorithms (MUSIC\cite{schmidt1986multiple} and SRP-PHAT\cite{dibiase2001robust}), each method is tested in section \ref{sec:Results} using the exact same testing datasets. These testing datasets consist in 3600 sources, randomly distributed in the torus depicted on Fig~\ref{fig:Pos_src}, emitting unseen signals during the training phase. In order to evaluate the robustness of each methods to noisy measurements in reverberating environments, the statistics of the DoA mismatch are derived from these 3600 positions, for each proposed propagating environments, and for different deterministic (SNR) values, ranging from $+40$~dB to $-1$~dB.\\

\subsection{\label{subsec:eval}Evaluation metrics}

In the present paper, the BeamLearning approach is evaluated both in a classification framework and in a regression framework. As a consequence, in order to analyze precisely the performances of the proposed DNN for the task of 2D-DoA SSL, several evaluation metrics will be used in the following. For the task of supervised multi-class angular classification, the overall accuracy is computed during the training phase and for testing using the number of correctly recognized class examples (true positives, $t_{p_i}$ ), the number of correctly recognized examples that do not belong to the class (true negatives, $t_{n_i}$ ), and examples that either were incorrectly assigned to the class (false positives, $f_{p_i}$) or that were not recognized as class examples (false negatives, $f_{n_i}$)\cite{sokolova2009systematic}. Other angular mismatch estimators will also be detailed in section \ref{subsec:number_classes}. For the task of supervised angular regression, the DoA mismatch between the estimated angular positions $\tilde{\theta(k)}$ and the groundtruth angular positions $\theta(k)$ is evaluated for each sources using statistics derived from the absolute angular mismatch $\vert  \tilde{\theta(k)} - \theta(k)\vert$ over all the sources in the testing dataset.

\section{BeamLearning network analysis \label{sec:Output}}


\subsection{\label{subsec:filterbank_analysis}Learnable filterbanks inner working analysis}

In this subsection, we analyse the variables learnt in the first subnetwork described in \ref{subsec:filterbank_subnetwork}, in order to give further insight on the learning process involved. The architecture of this first subnetwork of the BeamLearning DNN has been specifically developed to automatically build a 2D map $\mathbf{S}_{i,n}$ that can be interpreted as the filtering of raw input multichannel measurements by 128 different tunable filters. On contrary to conventional Beamforming, these 128 time domain outputs cannot be interpreted as beams, but rather as 128 optimized channels which aim at achieving a joint filtering in space and time. Since this subnetwork is composed of 3 filterbanks constituted by 6 depthwise separable atrous convolutions with a channel multiplier of 4 and a number of output channels of 128, it is important to note that the global output of this first subnet results from the combination of nonlinear filtering achieved by the equivalent $3 \times 6 \times 4 \times 128 = 9216 $ learnable filters followed by mathematical operations operated by the LN layer, and their nonlinear activation functions. Rather than looking at each of these filters independantly, this analysis aims at giving insightful representations of the spatio-temporal filtering achieved when the multichannel input data has partially passed through the $n$ first layers of the learnable filterbank subnetwork. This analysis will allow to better understand the multiresolution offered by the residual networks of one-dimensional depthwise separable atrous convolutions that constitute the main ingredient of this subnetwork.\\ 

\begin{figure*}[ht]
    \centering
  \subfloat[\label{fig:Directivite_1_1}]{%
       \includegraphics[width=0.333\linewidth]{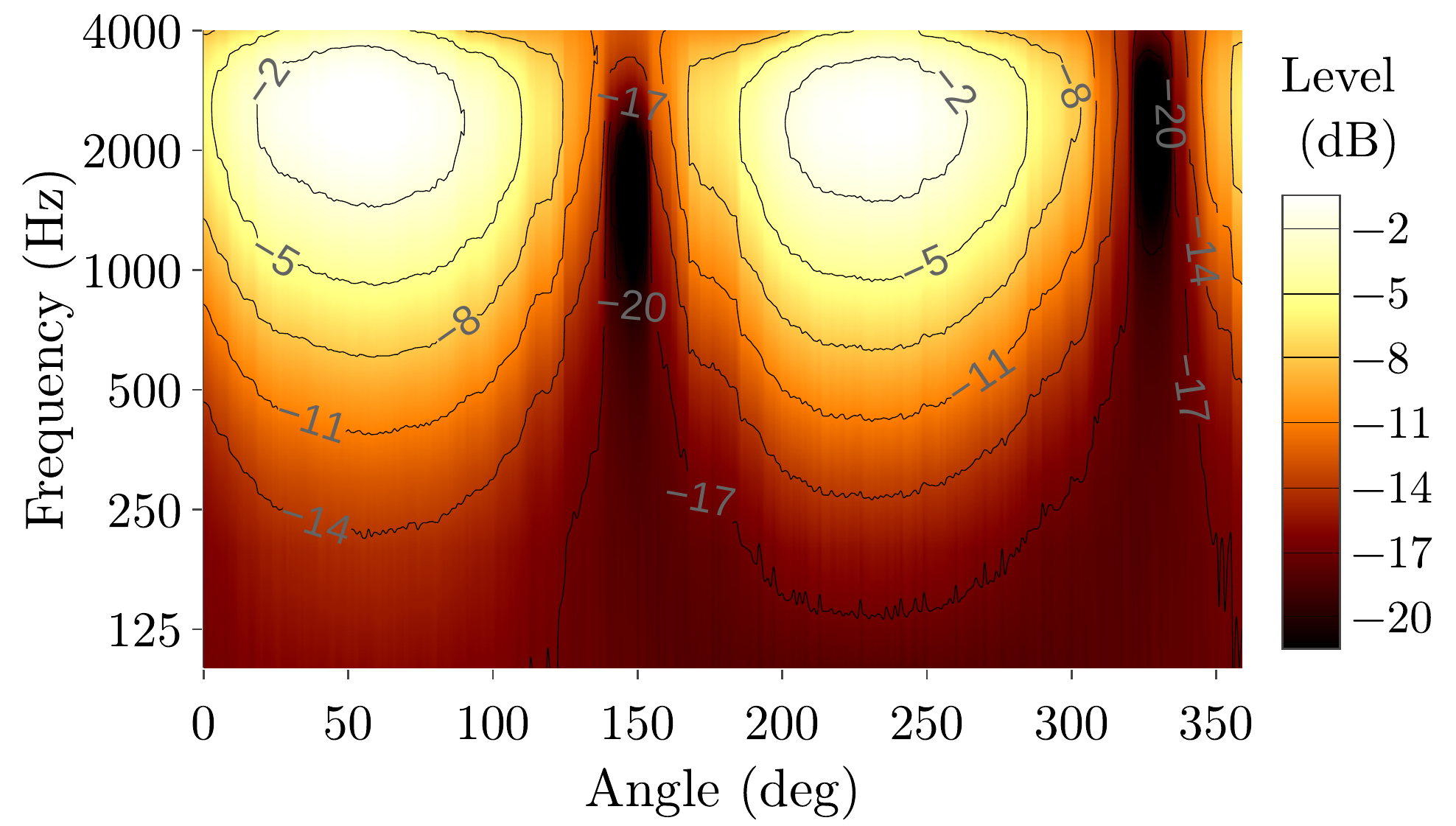}}
    \hfill
  \subfloat[\label{fig:Directivite_1_8}]{%
       \includegraphics[width=0.333\linewidth]{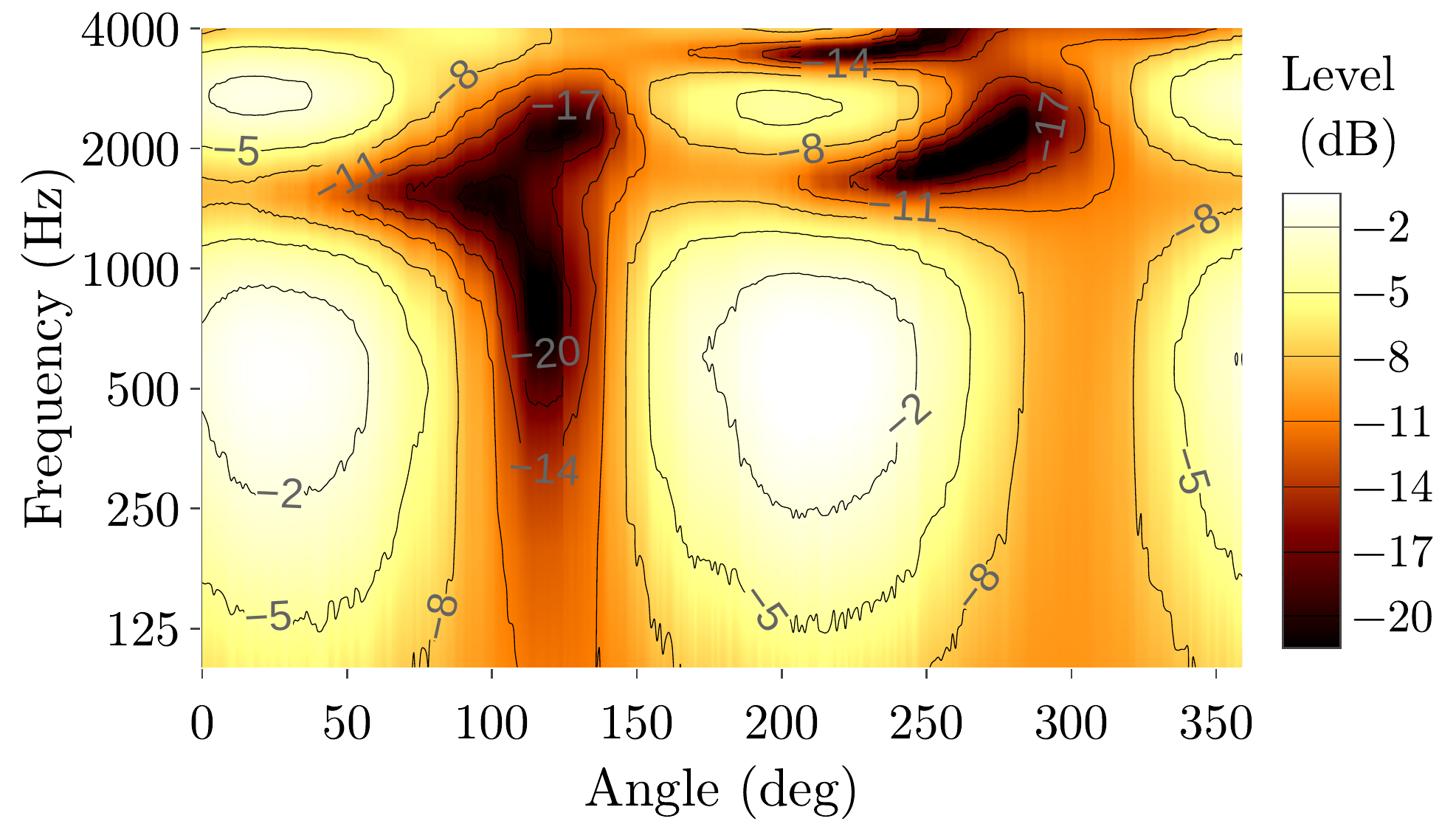}}
    \hfill
  \subfloat[\label{fig:Directivite_1_32}]{%
       \includegraphics[width=0.333\linewidth]{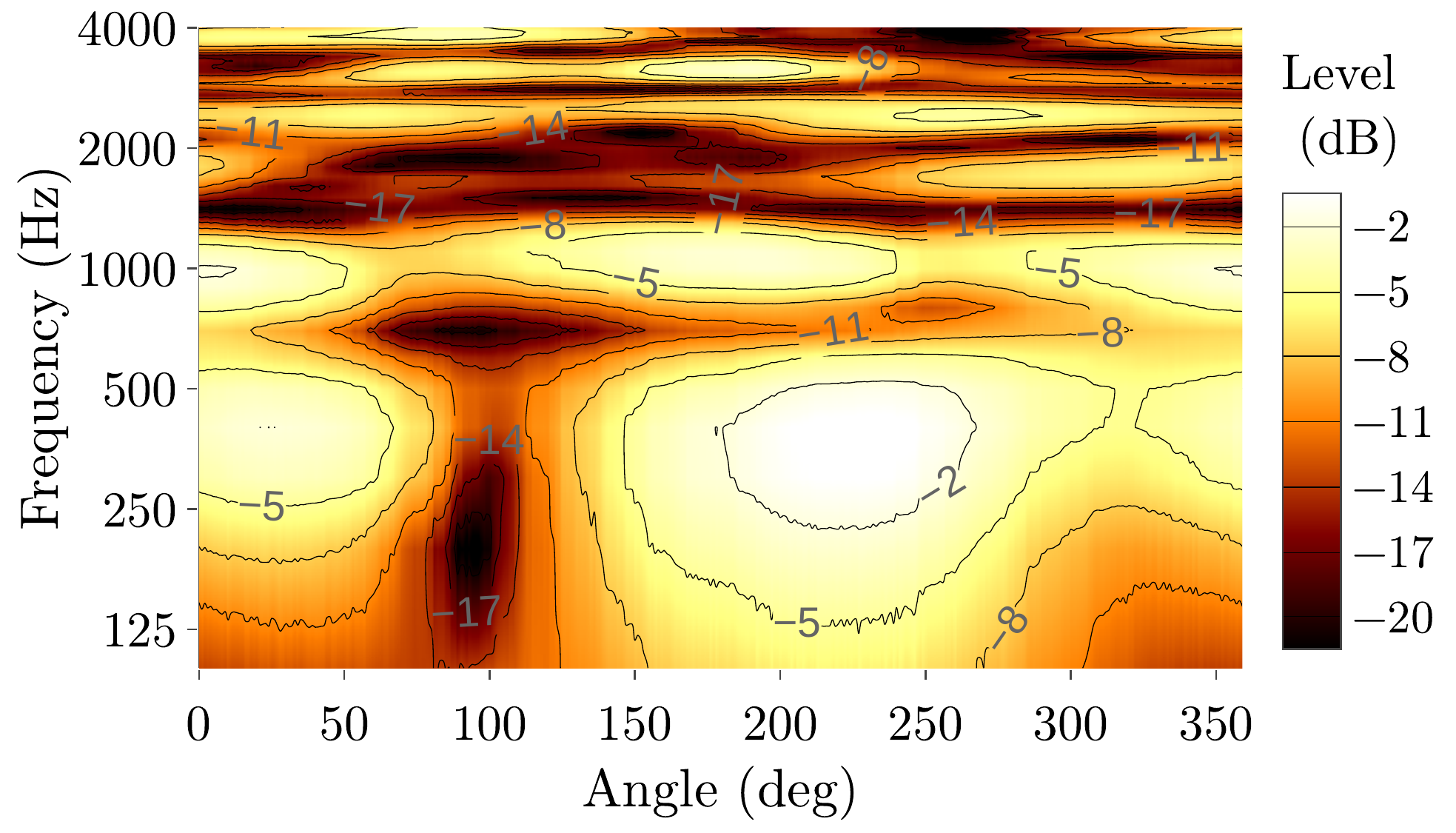}}
\caption{\label{fig:Directivite}Directivity maps (in dB) of one of the 128 equivalent filters obtained by passing through the first layers of the network, for the 1\textsuperscript{st} (a), 4\textsuperscript{th} (b), and 6\textsuperscript{th} (c) depthwise-separable atrous convolutional layers of the filterbank subnetwork. These 3 layers correspond respectively to a dilatation factor of 1, 8 and 32.}
\end{figure*}

In order to perform this analysis, the BeamLearning network is trained for a regression problem using the synthetic dataset described in \ref{subsec:Datasets}. This training dataset, will be referred in the following as $\mathcal{D}_{\text{train}}^{\text{ Room1}}$. In this dataset, each of the sources located in a room of dimensions $7\times10\times3.7$~m and a reverberation time of 0.5~s. emit ambient speech noise signals, car horn signals, and symphonic music. The data augmentation procedure detailed  with added measurement noise detailed in \ref{subsec:Datasets} is used with a $SNR \geq 20$ dB. Following the procedure detailed in \ref{subsec:Datasets}, the 48000 sources positions included in $\mathcal{D}_{\text{train}}^{\text{ Room1}}$ are randomly distributed in a torus of section $ R \times (2 \Delta R) \times \sin(2 \Delta \phi)$ and of central radius $R=2$~m, and with $\Delta R = 0.5$~m and $\Delta \phi = 7$ degrees (see Fig~\ref{fig:Pos_src}). All the parameters of the BeamLearning network are then frozen after convergence of the learning process. This frozen BeamLearning network is then used to build 2 dimensional $(\theta,f)$ magnitude maps of the $128\times18$ spatio-temporal filterings offered at the output of the 18 successive depthwise-separable atrous convolutional layers used in the learnable filterbank subnetwork. Among these  2304 possible $(\theta,f)$ transfer function magnitude maps offered at increasing depths of the subnetwork, Fig.~\ref{fig:Directivite} shows three of them. These three maps are selected from the first filterbank, and corresponding to dilatation factors of 1, 8, and 32 in order to illustrate the multiresolution capabilities offered by the BeamLearning approach.\\

 The representations in Fig.~\ref{fig:Directivite} are obtained using unseen input data during the DNN training. These unseen testing input data correspond to raw measurements on the microphone array using 360 sources uniformly sampled on a circle of radius 2~m at $\phi = \pi/2$, emitting monochromatic signals ranging from 100~Hz to 4000~Hz. This allows to compute the magnitude of the filtering process for each frequency and for each azimuthal angle at any of the layer composing the learnable filterbank subnetwork (including nonlinear activations and normalization layers), which is a strict equivalent of representing the directivity maps obtained at different depths of the subnetwork.\\
 
 When analysing the obtained directivity maps on Fig.~\ref{fig:Directivite}, one can observe that the frequency and spatial selectivity of the learnt filters increase with the depth through which the input data passes into the DNN. For the very first layer of the filterbank subnetwork corresponding to Fig.~\ref{fig:Directivite_1_1}, the obtained directivity obtained for this particular filter corresponds to a dipolar response above 1000~Hz, with directivity notches at $\theta = 150$~degree and $\theta = 330$~degree. At low frequencies, this particular filter (out of the 128 possible filters for this layer) exhibits a global attenuation of -20~dB, without any strong directivity pattern. It is also interesting to note that among the 128 filters obtained for this particular first convolutional layer (data not shown), most filters exhibit such a dipolar angular response at high frequencies, with various spatial notches values. This behaviour can be explained by the fact that for this layer, the filter kernels have a very small size of 3 samples, with a dilation factor of 1. \\
 
 However, when the data passes through more depthwise-separable atrous convolutional layers in the filterbank (Fig.~\ref{fig:Directivite_1_8} and Fig.~\ref{fig:Directivite_1_32}), the combination of optimized filters begin to be more and more selective, both in the frequency domain and in the angular domain. This is even observed at low frequencies, thanks to the kernel widthening offered by the increased dilation factors and the combination of filters at previous layers offered by the residual connections. This observation suggests that the stacking of the layers composing the three successive filterbanks converge to a global nonlinear filtering process into 128 channels, where each output channel specializes in several frequency bands and several angle incidences. When the source signal exhibits a particular spectrum in a reverberating environment, the energies carried by these 128 channels are then combined by the full connected layer in order to infer the source position.

\subsection{\label{subsec:classif_directivity} BeamLearning output analysis for a classification problem}

Most published studies on SSL using a deep learning approach treat the localization problem as an angular classification problem, where the objective is to determine to which discrete portion of the angular space the emitting source belongs to\cite{perotin2019crnn, chakrabarty2017broadband, he2018deep, takeda2016sound}. When seeking a better angular resolution for the SSL problem, these approaches require to increase the number of classes. \\

In this n-class classification approach, the source angular position $\theta \in [-180,180]$  is first converted to an integer $i$ representing the belonging of the source to the i\textsuperscript{th} angular partition among the n possibilities : 

$$i = \lfloor (\theta + 180) \frac{n}{360} \rfloor $$ 

Using this approach, $\theta$ is then approximated to a corresponding discrete $\theta_i$ (the center of the i\textsuperscript{th} angular partition), which is equivalent at discretizing the space on an angular grid : 

$$\theta_i = (i+\frac{1}{2}) \times \frac{360}{n} - 180   $$

The label used for classification can then be converted using a one-hot encoded target strategy using the integer $i$\cite{perotin2019regression,chakrabarty2017broadband}. Some authors also proposed the use of a soft Gibbs target\cite{perotin2019regression,he2018deep} exploiting the absolute angular distance between the true angular position $\theta$ and the attributed discrete value $\theta_i$ on the grid.\\

\begin{figure}[ht]
    \centering
  \subfloat[]{%
       \includegraphics[width=0.333\linewidth]{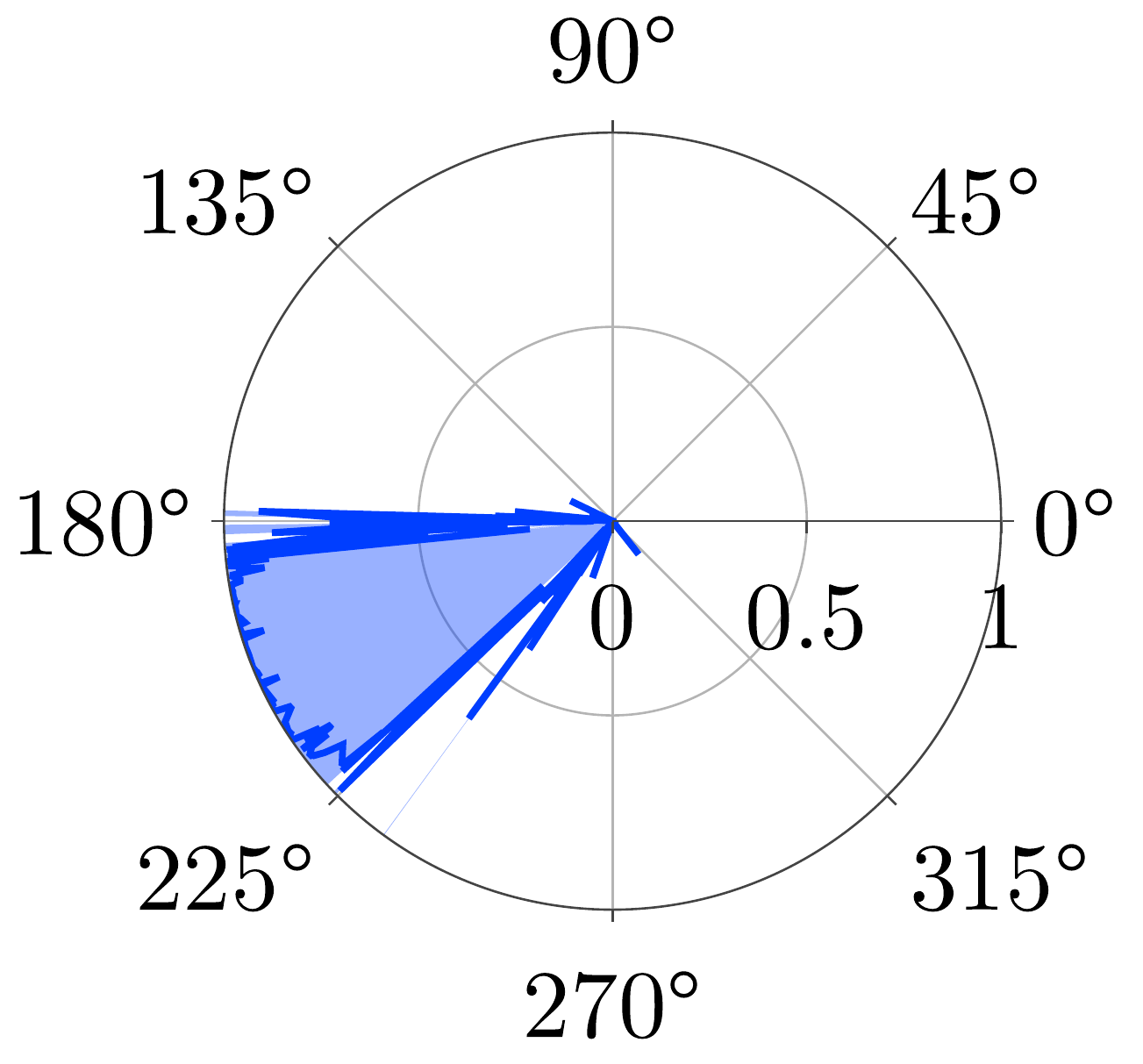}}
    \hfill
  \subfloat[]{%
       \includegraphics[width=0.333\linewidth]{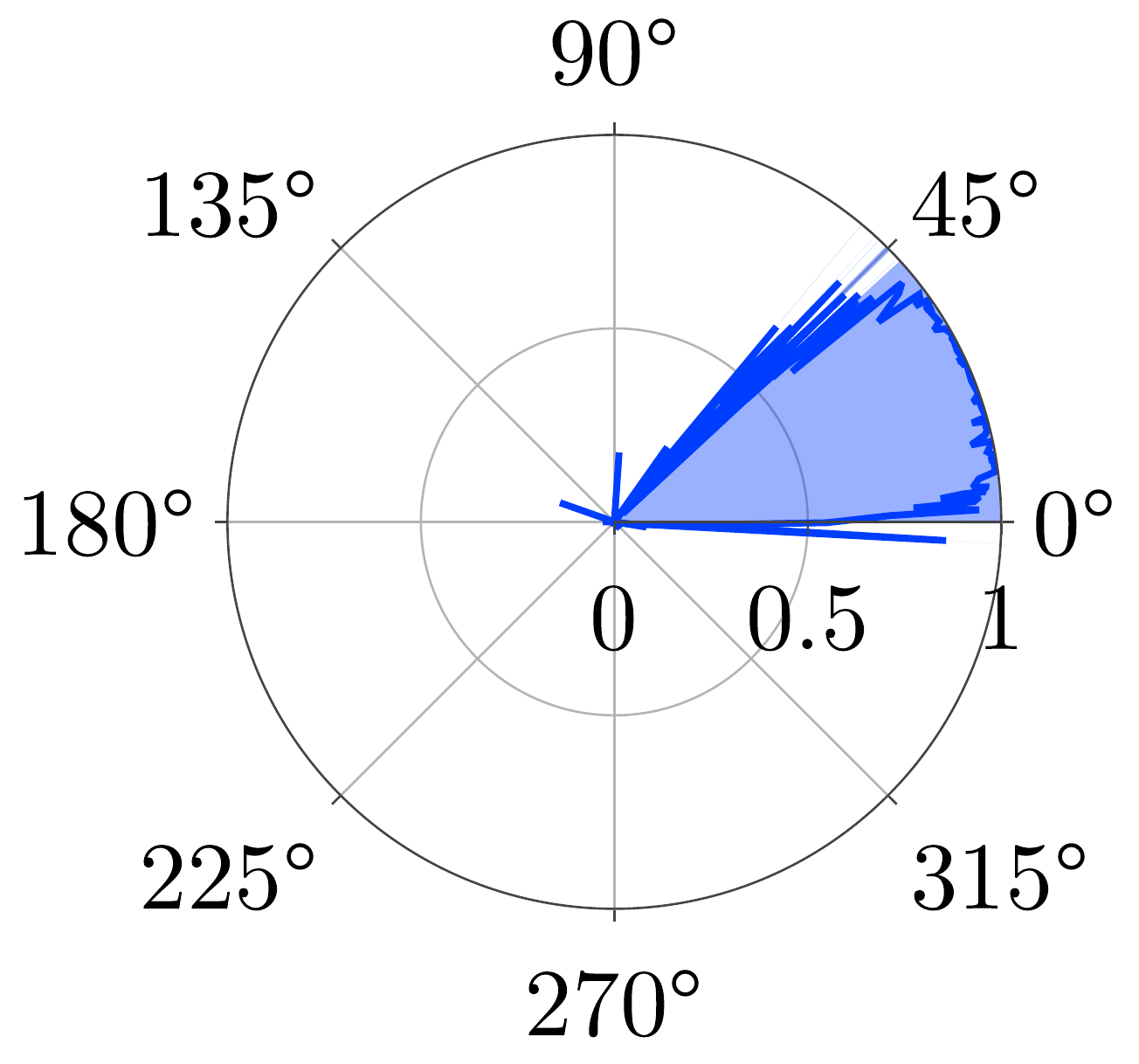}}
    \hfill
  \subfloat[]{%
       \includegraphics[width=0.333\linewidth]{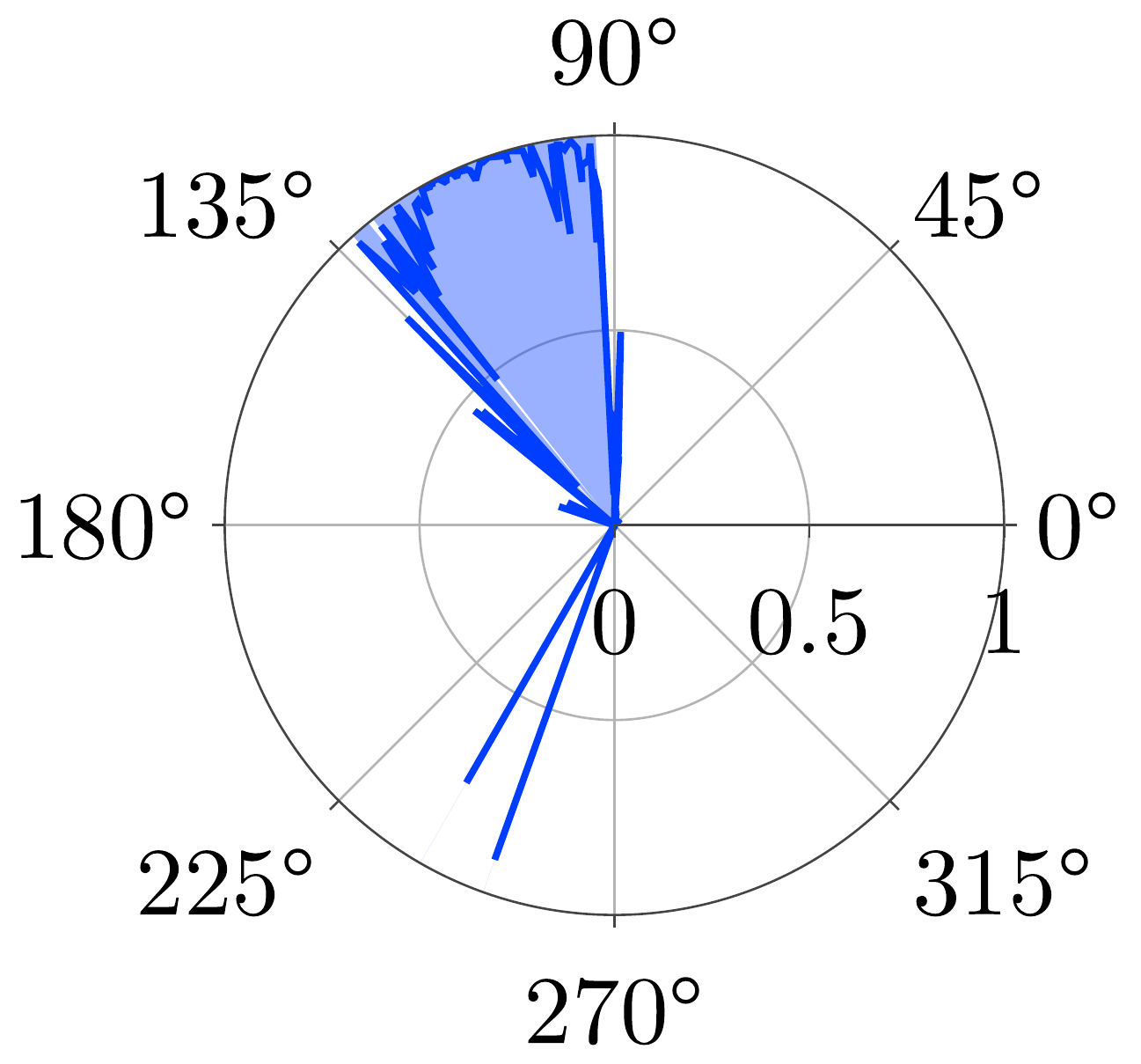}}
\caption{\label{fig:Directivite_classif}{Directivity diagrams of 3 of the 8 output neurons for the SSL task seen as a classification problem. The diagrams are obtained by plotting the output probability (solid blue) obtained for each output neuron when testing the BeamLearning network with the 4800 sources positions in $\mathcal{D}_{\text{test speech}}^{\text{Room1}}$. The sources positions where the predicted output is the i\textsuperscript{th} angular partition are depicted using light blue shaded areas.}}
\end{figure}

In the following, we study the behaviour of the DNN output neurons for a 8-class classification problem. The analysis of the BeamLearning DNN outputs for a regression problem will be carried out similarily in \ref{subsec:regression_directivity}. For the classification problem, the BeamLearning DNN is trained with a cross-entropy loss function preceded by a softmax activation function and with one-hot encoded labels. Using the same training dataset $\mathcal{D}_{\text{train}}^{\text{ Room1}}$ as in \ref{subsec:filterbank_analysis} with data augmentation procedure with $SNR \geq 20$~dB and after convergence of the BeamLearning network, the frozen DNN is fed with unseen input data during the DNN training. These unseen testing inputs corresponds to 4800 sources randomly distributed over a torus of mean radius $R=2$~m, emitting unseen speech signals. This testing dataset, which will be used several times throughout the present paper, will be referred in the following as $\mathcal{D}_{\text{test speech}}^{\text{Room1}}$. The use of such a testing dataset allows to represent quantitatively the probability value obtained using the softmax function applied to the estimated multiclass label for each of the 8 directions as a polar representation, which can be interpreted as a directivity diagram (see Fig.\ref{fig:Directivite_classif}).\\

 The analysis of the obtained directivity diagrams on Fig.~\ref{fig:Directivite_classif} allows to highlight the fact that during learning, each output neuron specializes to its corresponding angular partition, with a sharp directivity pattern. However, a precise analysis of these directivity pattern shows that some of these directivity patterns slightly overlap adjacent angular partitions that define the angular classes used for the SSL problem. As a consequence, a loss of accuracy can be observed for sound sources located in the vicinity of angular partitions. This also clearly shows that the misclassified sources mostly belong to contiguous angular sectors to the estimated angular sector. This suggest that for a rather coarse angular grid, the use of a soft Gibbs target or a Gibbs-weighted cross-entropy loss such as proposed in\cite{perotin2019regression,he2018deep} may only offer small accuracy improvements for the proposed BeamLearning approach over the simple one-hot encoded labelling strategy we use. However, for finer angular grids and with a higher number of angular partitions, the number of boundaries between classes increase, which could lead to a loss of global classification accuracy, which may justify the use of a regression approach.     
 
\subsection{\label{subsec:number_classes}Influence of the number of angular classes}

Since an angular classification problem is directly linked to the choice of a discrete angular grid to define angular partitions, the present subsection aims at giving further insight on the influence of the number of classes when aiming at obtaining a better angular resolution for the SSL problem. As observed in the previous section, for a small number of classes, most misclassified source positions are located in the vicinity of the chosen angular partitions. In order to verify if this assumption still stands for a high count of classes, we conducted several n-class classification experiments, with $n$ ranging from 8 to 128. Those experiments have been conducted both with the experimental dataset $\mathcal{D}_{}^{\text{exp}}$ detailed in \ref{subsec:Datasets} with $\Delta R = 0.5~m$ and $\Delta \phi = 7^{\circ}$ partitionned in $\mathcal{D}_{\text{train}}^{\text{exp}}$ and $\mathcal{D}_{\text{test}}^{\text{exp}}$, and the same training and testing synthetic datasets $\mathcal{D}_{\text{train}}^{\text{ Room1}}$ and $\mathcal{D}_{\text{test speech}}^{\text{Room1}}$  than those used in \ref{subsec:filterbank_analysis}. Though, it is interesting to note that for the experimental dataset $\mathcal{D}_{}^{\text{exp}}$, the BeamLearning network has been trained using 72000 acoustic sources generated using the laboratory's higher order ambisonics sound field synthesis system\cite{lecomte2016fifty}, each source emitting a simple set of 6 pure sinusoidal signals, whose frequencies correspond to the central frequencies of the octave bands between 125 Hz and 4000 Hz. Since the sound field synthesis system is located in a rather moderately reverberating environment ($T_R \approx 0.2$~s) and since the frequency content and the dynamics of those emitted signals are much less diverse than in $\mathcal{D}_{\text{train}}^{\text{ Room1}}$, this represents an easier learning task than the one achieved using the synthetic dataset.\\

For both training datasets and for each number of classes $n$, after convergence of the learning process, the frozen Beamlearning DNNs are fed with unseen input data during the training phase. These unseen testing inputs corresponds to 4800 sources randomly distributed over a torus of mean radius $R=2$~m around the microphone array emitting unseen signals during the DNN training phase. Table~\ref{tab:Prec_classif} presents the obtained performances in terms of multi-class accuracy and in terms of angular accuracy. Since the SSL determination is treated here as a classification problem, two kinds of angular accuracies can be calculated, which both carry some insightful information about the localization performances of the algorithm :

$$ \Delta \theta_{class} = \overline{\vert \bm{\tilde{\theta_i}} - \bm{\theta_i} \vert} $$ 

$$ \Delta \theta = \overline{\vert \bm{\tilde{\theta_i}} - \bm{\theta} \vert} $$

, where $\bm{\theta}$ is the set of continuous ground truth positions of the 4800 testing sources, $\bm{\theta_i}$ is the set of corresponding groundtruth grid positions defined in \ref{subsec:classif_directivity}, and $\bm{\tilde{\theta_i}}$ is the set of discrete grid positions estimated by the BeamLearning network.\\

\renewcommand{\arraystretch}{1.5}

\begin{table}[ht]
	\caption{Multiclass classification performances (accuracy, and mean absolute angular errors) with increasing number $n$ of classes. Results are presented both for an experimental dataset and a synthetic dataset in a reverberating environment, with a $SNR = 20$~dB. The testing dataset is composed of 4800 positions randomly distributed around the microphone array. Results with both $\Delta \theta \leq 5^{\circ}$ and $\frac{\Delta \theta}{\alpha} \leq 1$ are in bold.}
		\begin{tabularx}{\linewidth}{|X|X||X|X|X|X|X|}
		    \hline\hline
			\multicolumn{2}{|c||}{$n$} & 8 & 16 & 32 & 64 & 128  \\ 
			\hline
			\multicolumn{2}{|c||}{Half angular width $\alpha$} & $22.5^{\circ}$ & $11.25^{\circ}$ & $5.63^{\circ}$ & $2.81^{\circ}$ & $1.4^{\circ}$  \\ 

			\hline 
			\hline 
			& Accuracy & 99.2\% & 97.1\%& \textbf{94.3\%} & \textbf{91.3\%} & \textbf{87.7\%}   \\  
			\raisebox{-.5\normalbaselineskip}[0pt][0pt]{\rotatebox[origin=c]{90}{\parbox{0.8cm}{\centering{Experim.  dataset}}}} & $\Delta \theta_{class}$  & $0.3^{\circ}$ & $0.6^{\circ}$ & $\mathbf{0.6^{\circ}}$ & $\mathbf{0.5^{\circ}}$ & $\mathbf{0.4^{\circ}}$  \\ 
			& $\Delta \theta$  & $11.1^{\circ}$ & $5.8^{\circ}$ & $\mathbf{3.0^{\circ}}$ & $\mathbf{1.5^{\circ}}$ &  $\mathbf{0.9^{\circ}}$  \\ 
\hline
			& Accuracy  & 92.1\% & 87.6\%& \textbf{79.5\%} & 61.4\% & 37.9\%  \\ 
			\raisebox{-.5\normalbaselineskip}[0pt][0pt]{\rotatebox[origin=c]{90}{\parbox{0.8cm}{\centering{Synthetic  dataset}}}}  & $\Delta \theta_{class}$  & $5.9^{\circ}$ & $5.1^{\circ}$ & $\mathbf{3.0^{\circ}}$ & $4.1^{\circ}$ & $4.7^{\circ}$   \\ 
			& $\Delta \theta$  & $14.9^{\circ}$ & $8.8^{\circ}$ & $\mathbf{4.3^{\circ}}$ & $4.5^{\circ}$ &  $4.8^{\circ}$   \\ 
    \hline\hline
		\end{tabularx} 
	\label{tab:Prec_classif} 
\end{table}

The analysis of the results in Table~\ref{tab:Prec_classif} allow to show that even with a rather low classification accuracy for a high count of angular classes, the mean absolute angular errors on the grid $\Delta \theta_{class}$ remain very low and almost constant across the number of classes. This suggests that for a high count of angular classes, the growing number misclassified sources are attributed to grid points that remain close to their ground truth positions. Since the angular grid refines when $n$ increases, the observed decrease of classification accuracy is therefore compensated by the increasing of the angular resolution offered by a higher count of angular classes. However, there is a clear influence of this grid resolution when estimating $\Delta \theta$, which also includes the uncertainty due to the grid resolution. Indeed, the computation of $\Delta \theta$ also includes the absolute angular error for sources that are classified in the right angular partition. One can observe that for the synthetic dataset in a reverberating environment, $\Delta \theta$ remains smaller than the half angular width $\alpha$ of the angular partitions for $n \leq 32$. However, for $n=64$ and $n=128$, the mean absolute angular error $\Delta \theta$ exceed $\alpha$, which signifies that for such a angular resolution, a regression approach could be more appropriate.\\

\begin{table}[ht]
	\caption{Statistics over the misclassified populations $\Omega_m$ with increasing number of classes $n$ (percentage of misclassified sources , and mean absolute angular errors) with increasing number $n$ of classes. Results are presented both for an experimental dataset and a synthetic dataset in a reverberating environment, with a $SNR = 20$~dB. The testing dataset is composed of 4800 positions randomly distributed around the microphone array. Results with both $P_{adj} \geq 85 \%$ and $\frac{\beta}{\alpha} \leq 1$ are in bold.}
		\begin{tabularx}{\linewidth}{|X|X||X|X|X|X|X|}
				    \hline\hline
			\multicolumn{2}{|c||}{$n$} & 8 & 16 & 32 & 64 & 128  \\ 
			\hline
			\multicolumn{2}{|c||}{Half angular width $\alpha$} & $22.5^{\circ}$ & $11.25^{\circ}$ & $5.63^{\circ}$ & $2.81^{\circ}$ & $1.4^{\circ}$  \\ 

			\hline 
			\hline 
			& Card($\Omega_m$) & \textbf{34}  & \textbf{139} & \textbf{273} & \textbf{417} & \textbf{591} \\  
			\raisebox{-.5\normalbaselineskip}[0pt][0pt]{\rotatebox[origin=c]{90}{\parbox{0.8cm}{\centering{Experim.  dataset}}}} & $P_{adj}$  & $ \mathbf{100 \%}$ & $ \mathbf{100 \%}$ &$ \mathbf{100 \%}$ &$ \mathbf{100 \%}$ & $\mathbf{100 \%}$   \\ 
			& $\beta$  & $\mathbf{0.01^{\circ}}$ & $\mathbf{0.03^{\circ}}$ & $\mathbf{0.09^{\circ}}$ & $\mathbf{0.17^{\circ}}$ &  $\mathbf{0.37^{\circ}}$  \\ 
\hline
			& Card($\Omega_m$) & \textbf{379} & \textbf{595} & \textbf{984}  & \textbf{1853} & 2981   \\  
			\raisebox{-.5\normalbaselineskip}[0pt][0pt]{\rotatebox[origin=c]{90}{\parbox{0.8cm}{\centering{Synthetic  dataset}}}} & $P_{adj}$  & $ \mathbf{88.9 \%} $ & $ \mathbf{82.2 \%} $ & $ \mathbf{94.6 \%}$ & $ \mathbf{88.1 \%} $ & $ 57.8 \%$  \\ 
			& $\beta$  & $\mathbf{1.8^{\circ}}$ & $\mathbf{2.0^{\circ}}$ & $\mathbf{1.0^{\circ}}$ & $\mathbf{2.4^{\circ}}$ &  $3.0^{\circ}$  \\ 
				    \hline\hline
		\end{tabularx}
	\label{tab:Stats_errors_classif} 
\end{table}

It is also interesting to study the spatial repartition of the misclassified sources. Table~\ref{tab:Stats_errors_classif} presents two insightful statistics over the population $\Omega_{m}$ of misclassified sources. $P_{adj}$ represents the percentage of $\Omega_{m}$ elements whose groundtruth position is located in an contiguous angular partition to the estimated angular partition. $\beta$ represents the mean absolute angular distance between the ground truth position of  misclassified sources and the boundary of the mis-estimated angular partition.\\

The analysis of these metrics -- and in particular of $P_{adj}$ -- allows to conclude that even with a high count of classes and a refined discrete grid of possible angles, the great majority of misclassified sources mostly belong to contiguous angular sectors to the estimated angular sector. The only situation where misclassified sources begin to spread on non contiguous angular partitions corresponds to a 128-class classification problem in a reverberating and noisy environment. Using the proposed BeamLearning network, this observation again advocates for the limited interest of the use of a proximity-aware Gibbs weighing strategy proposed in\cite{perotin2019regression,he2018deep} for labelling the sound sources positions. The proposed BeamLearning network is only trained for classification with one-hot encoded labels that do not take into account any proximity weighting. This result suggests that the obtained representation at the output of the learnable filterbanks and the pseudo-energy network already encodes efficiently this information. \\

Furthermore, the analysis of  $\beta$ shows that even with a high count of classes, the misclassified sources even remain very close to the boundary of the estimated angular section (less than $3^{\circ}$ in all cases) . Except for  $n=64$ and $n=128$ classification problems in a reverberating and noisy environment, the mean distance to the boundary remains smaller than the quarter width of the angular sectors. However, when seeking an angular resolution of less than $5^{\circ}$ (corresponding to $n > 32$), these results suggest that a regression approach could be more appropriate and less dependent to the choice of the angular partitions, which confirms the results obtained in\cite{perotin2019regression} for a similar problem in 3D, but with a completely different DNN architecture and different kind of input signals representation.

\subsection{\label{subsec:regression_directivity} BeamLearning output analysis for a regression problem}

\begin{figure}[ht]
	\centering
	\includegraphics[width=0.75\columnwidth]{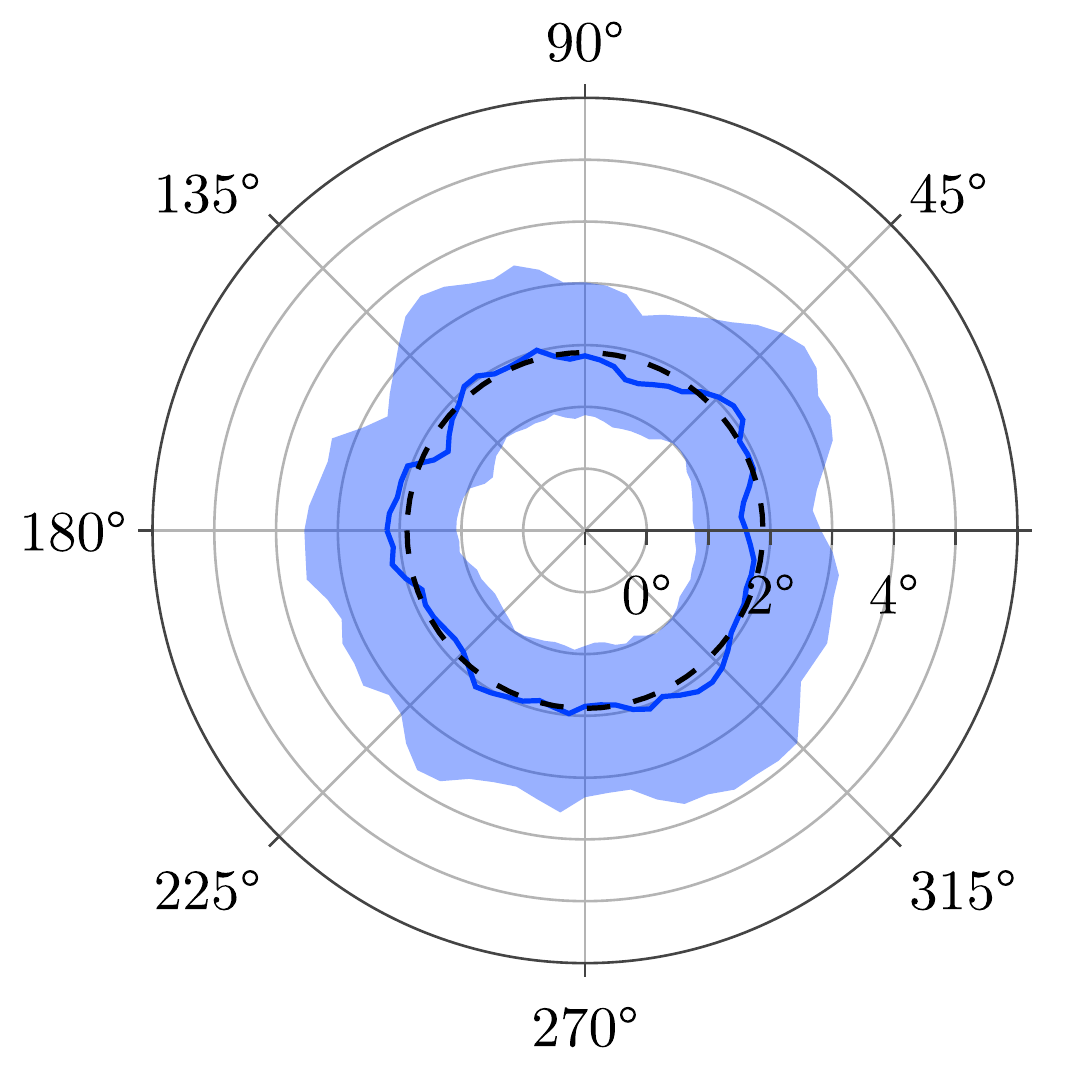}
	\caption{\label{fig:Err_circ} Angular diagram of statistical performance metrics derived from the absolute angular mismatches obtained by the BeamLearning DNN using an unseen testing dataset $\mathcal{D}_{\text{test speech}}^{\text{Room1}}$ with a fixed $SNR = 20$~dB. Global median over the whole dataset (dashed black line), local sliding median over an angular window of 5 degrees (solid blue line), and local interquartile range (light blue shaded area).}
\end{figure}

In this subsection, the BeamLearning DNN is used for DoA determination, in a regression scenario. As explained in sections \ref{subsec:output} and \ref{subsec:trainig}, in  this case, the sound source angular position is inferred from the output vector of the network, which represents the estimated cartesian coordinates of the unit vector pointing to the source DoA. The whole BeamLearning DNN architecture remains unchanged with respect to the classification problem, except for the last layer and the cost function described in section \ref{subsec:trainig}. In a regression framework though, the DNN output cannot be used to derive directivity patterns as it was in \ref{subsec:classif_directivity}, since the output is directly linked to the continuous DoA space. However, a polar representation of the absolute angular errors committed by the DNN over a large testing dataset (see Fig.~\ref{fig:Err_circ}) allows to represent the SSL performances of the proposed approach in a compact way.\\

The polar diagram on Fig.~\ref{fig:Err_circ} has been obtained by freezing the BeamLearning network after convergence of the training process for a DoA regression task in a reverberating environment, trained with a variable $SNR \geq 20$ dB. The training dataset used for this task is the same synthetic dataset $\mathcal{D}_{\text{train}}^{\text{ Room1}}$ as in \ref{subsec:filterbank_analysis}, \ref{subsec:classif_directivity}, and \ref{subsec:number_classes}. The frozen BeamLearning DNN is then fed with input data $\mathcal{D}_{\text{test speech}}^{\text{Room1}}$ (unseen during the training phase), with a fixed SNR~$= 20$~dB. For each of the 4800 testing positions, 50 different draws of random noise have been added to the microphone measurements. Using the $4800\times50$ obtained DoA estimations, statistics of the absolute angular errors are computed, including the global median over $\mathcal{D}_{\text{test speech}}^{\text{Room1}}$ ($1.9^{\circ}$), the local median computed using a sliding average over a window of 5 degrees (ranging from $1.5^{\circ}$ to $2.2^{\circ}$), and the local interquartile range (ranging from $0.7^{\circ}$ to $3.9^{\circ}$) which represents the confidence interval of the DoA mismatch estimations. Since this experiment uses the same training and testing datasets as those used for classification experiments in previous subsections, this allows to compare the obtained performances for this particular 2D DoA task. The obtained results show that the proposed regression approach with the BeamLearning DNN is more accurate than any of the classification test cases, even with a high count of classes. As a consequence, in the following, all the results and experiments will be conducted using a regression framework.

\section{Results and discussion using a regression approach \label{sec:Results}}

In this section, the SSL performances of the BeamLearning approach will be extensively studied and systematically compared with the performances of two state-of-the-art model-based SSL algorithms (MUSIC and SRP-PHAT) for a 2D DoA determination problem. This choice has been made using extensive preliminary comparisons between the TOPS, CSSM, wideband MUSIC, SRP-PHAT, and FRIDA algorithms included in the Pyroomacoustics library\cite{pyroomacoustics} and by selecting the most robust methods among them. \\

SRP-PHAT\cite{dibiase2001robust} is a widely used algorithm, which is often considered by the scientific community as one of the ideal strategies for SSL systems\cite{do2010srp,badali2009evaluating,cobos2010modified}. This algorithm aims at increasing robustness to signal and ambient conditions by combining the robustness of the steered response  power\cite{dmochowski2010steered} (SRP) approach with phase transformation filtering\cite{zhang2008does} (PHAT).\\

 The MUSIC algorithm\cite{schmidt1986multiple} and its wideband extensions \cite{argentieri2007broadband} is also a widely used and tested algorithm\cite{argentieri2007broadband,dmochowski2007broadband,ishi2009evaluation,zhang2009robust} for DoA estimation in noisy and reverberant environments, which also makes it an excellent choice for comparative tests. This algorithm, which is able to handle arbitrary geometries and multiple simultaneous sources, belongs to the family of subspace approaches, and depends on the eigen-decomposition of the covariance matrix of the observation vectors. In the present paper, we used the broadband implementation  of the MUSIC algorithm proposed in \cite{pyroomacoustics,argentieri2007broadband}, which is obtained by decomposing the wideband signal into multiple frequency bins, and then applying the narrowband algorithm for each frequency bin in order to get the source direction. In the following, for each methods, frequency content within $[100~Hz;4~kHz]$ is used for DOA estimations. This frequency range corresponds approximately to 90 frequency bins for the wideband MUSIC and SRP-PHAT methods.\\
 
 It is also important to note that the MUSIC and SRP-PHAT methods are both based on an angular grid search. Therefore, we used a 1-degree angular grid resolution, which is smaller than the observed interquartile ranges of the three compared SSL methods. Although this affects the computational cost of the two traditional DOA estimation algorithms, it ensures a fair comparison of DOA estimation performance between BeamLearning, MUSIC, and SRP-PHAT.\\
 
 In the following, a particular attention will be paid to the robustness to measurement noise in reverberating environments, to changes in the propagation environment or in the characteristics of the emitted signals. The computational efficiency of the three localization methods will also be compared for realtime 2D-DoA determination. 

\subsection{Robustness to noisy measurements \label{subsec:Robustness_noise}}

In this subsection, we evaluate the SSL performances of the BeamLearning DNN along with those of the MUSIC and SRP-PHAT algorithms, which both have been developed to accomodate to noisy measurement conditions, ranging from sensor ego-noise and acquisition boards electronic noise, to residual acoustic noise due to other interference sources. As explained in \ref{subsec:Datasets}, during the training phase, each multichannel recordings of the training mini batches are added to a random multichannel Gaussian noise with randomly picked signal-to-noise ratio (SNR) values in the range of $[x,+\infty[$~dB. This data augmentation procedure intends at offering an improved robustness to noisy measurements for SSL.\\

\begin{figure}[ht]
	\includegraphics[width=\columnwidth]{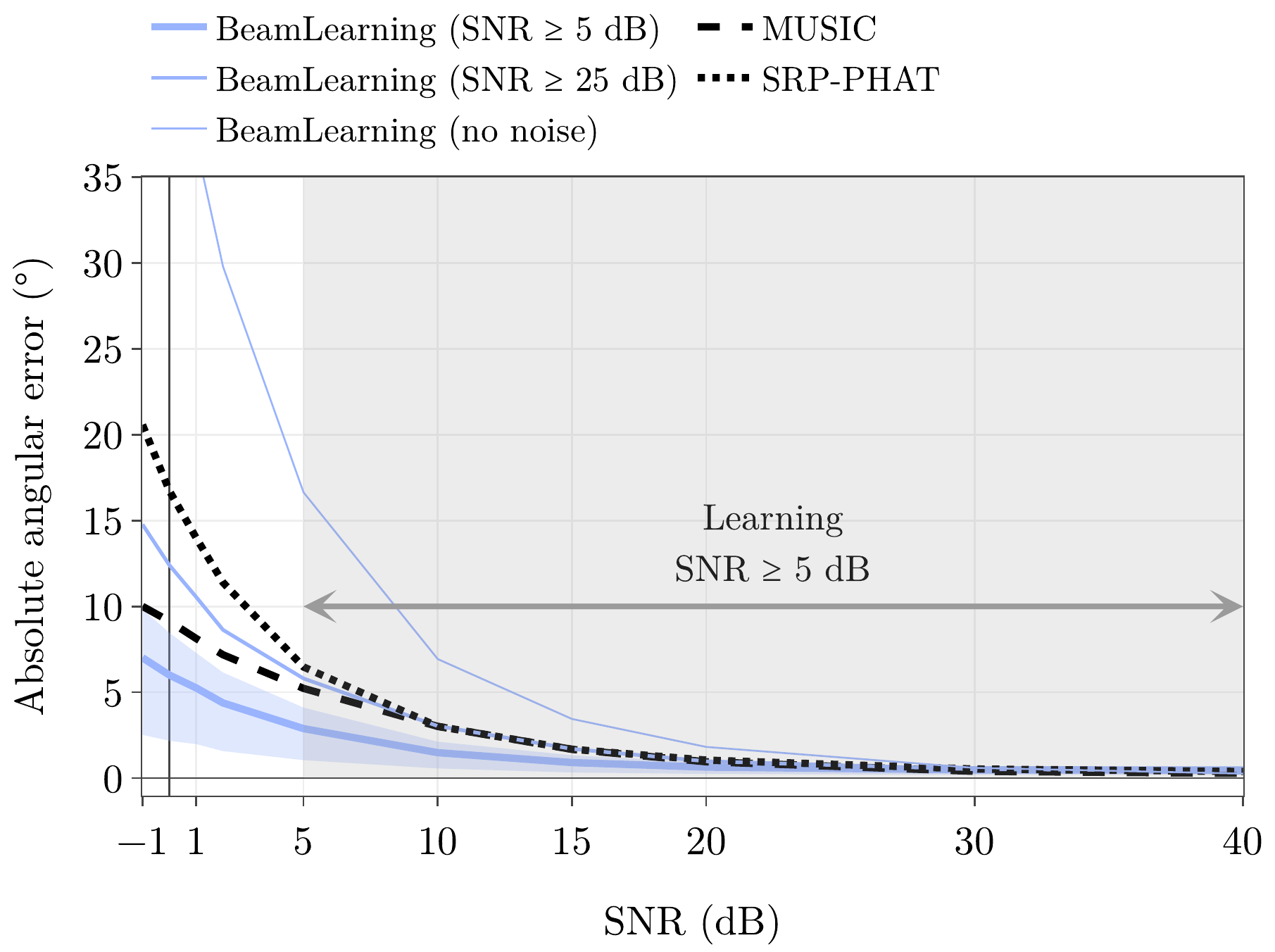}
	\caption{\label{fig:Libre}{(Color online) Mean absolute angular mismatch obtained in an anechoïc environment for different SNRs, using an unseen speech testing dataset (3600 positions for each SNR). The DoA errors curves are plotted for wideband MUSIC (dashed), SRP-PHAT  (dotted), and the proposed BeamLearning approach (solid blue lines). In order to emphasize the influence of the proposed data augmentation strategy, the BeamLearning DNN has been pre-trained three different times using noisy measurements, with random SNR~$\geq$~5~dB (thick blue line), with random SNR~$\geq$~25~dB (medium blue line), and without any noise added (thin blue line). The interquartile range of absolute angular mismatch is also represented for BeamLearning trained with SNR~$\geq$~5~dB (light blue shaded area).}}
	
\end{figure}

The BeamLearning DNN has been trained and frozen after convergence for two different environments: a perfectly anechoïc environment, and the same reverberating room than in previous subsections. In the anechoïc environment, the training procedure has been performed  with three different minimum SNR values ($x = 5$~dB, $x = 25$~dB, and $x = +\infty$~dB, \emph{i.e.} noiseless data) for data augmentation. In the reverberating environment, results are presented with a data augmentation procedure with $x = 5$~dB. This allows to observe the robustness to measurement noise only (see Fig.~\ref{fig:Libre}), and to demonstrate the capabilities of the proposed approach in both noisy and reverberating environments (see Tab.~\ref{tab:Prec_loca_reverb}).\\

For both environments, the testing datasets used for the evaluation of the MUSIC and SRP-PHAT algorithms and the frozen BeamLearning DNNs are composed of noisy measurements of 3600 sources positions emitting speech signals at a distance of $2\pm0.5$~m from the microphone array,  with a varying SNR ranging from  -1~dB to $+\infty$~dB. Both the sources positions and the emitted signals were unseen during the training phases of the BeamLearning DNN.  The analysis of the obtained results presented on Fig.~\ref{fig:Libre} and Tab.~\ref{tab:Prec_loca_reverb} shows that the proposed BeamLearning approach is significantly more insensitive to measurement noise than MUSIC and SRP-PHAT, both in a free field situation and in presence of reverberation. \\

\newcolumntype{b}{X}
\newcolumntype{s}{>{\hsize=.4\hsize}X}

\begin{table}[ht]
	\caption{Mean absolute angular mismatch obtained in a reverberating environment (RT~=~0.5~s), for different SNRs, using an unseen speech testing dataset (3600 positions). The DNN has been pre-trained using noisy measurements, with random SNR~$\geq$~5~dB. The best results are in bold.}
		\begin{tabularx}{\linewidth}{|b||s|s|s|s|s|s|}
						    \hline\hline

			SNR (dB) & -1 & 2 & 5 & 10 & 15 & 20 \\ 
			\hline 
			SRP-PHAT & $41.6^{\circ}$ & $28.7^{\circ}$ & $17.4^{\circ}$  & $7.9^{\circ}$ & $4.6^{\circ}$ & $3.4^{\circ}$ \\ 
			\hline 
			MUSIC & $27.5^{\circ}$ & $21.3^{\circ}$ & $17.8^{\circ}$ & $13.1^{\circ}$ & $11.0^{\circ}$ & $10.9^{\circ}$ \\ 
			\hline 
			BeamLearning & $\bm{21.1}^{\circ}$ & $\bm{13.0}^{\circ}$ & $\bm{8.2}^{\circ}$ & $\bm{4.8}^{\circ}$ & $\bm{3.4}^{\circ}$ & $\bm{2.7}^{\circ}$ \\ 
 \hline\hline
		\end{tabularx}
		\label{tab:Prec_loca_reverb} 
\end{table}

A comparison of the results obtained with the BeamLearning approach on Fig.~\ref{fig:Libre} allows to conclude that the proposed data augmentation process offers a very efficient way to improve SSL estimations. Over all the proposed DOA estimation methods in Fig.~\ref{fig:Libre}, the best results are obtained using the BeamLearning approach trained using a data augmentation process with a random SNR~$\geq~5$~dB. This result still stands when the testing dataset is used with SNR as small as $-1$~dB. For a moderate noise data augmentation (a random SNR training larger than 25~dB), the BeamLearning approach offer very similar SSL performances than both MUSIC and SRP-PHAT for unseen testing data with a SNR~$\geq~10$~dB. Without any data augmentation, the BeamLearning approach only outperforms MUSIC and SRP-PHAT for unseen testing data with a SNR~$\geq~30$~dB. It is also interesting to note that the use of the proposed data augmentation mechanism does not degrade the localization accuracy at high SNR, even if the data augmentation uses very noisy data. This property is directly linked to the fact that the training process is performed with randomly picked signal-to-noise ratio (SNR) values in the range of $[x,+\infty[$ dB, which does not give more importance to noisy data than to noise-free data when training the BeamLearning DNN.\\

 In free-field (Fig.~\ref{fig:Libre}) and for low-noise measurements, the three methods offer very similar SSL performances. However, as expected, the MUSIC algorithm performs significantly better than SRP-PHAT for heavily noisy measurements (SNR~$\leq$~5~dB). It is although interesting to note that the proposed BeamLearning approach performs even better than MUSIC at low SNRs in free field. Since the mean absolute angular errors obtained using wideband MUSIC and SRP-PHAT stand outside the interquartile range of the BeamLearning approach trained with a random SNR~$\geq~5$~dB, one can also conclude that the improvement offered by the proposed BeamLearning approach is statistically significant over those two methods at low SNR : the DOA estimation precision is improved over MUSIC by 30\% (from 10.0\textsuperscript{o} to 7.0\textsuperscript{o} at -1~dB).\\
 
 The same trend is also observed in a reverberating environment (Tab.~\ref{tab:Prec_loca_reverb}), where the BeamLearning approach offers better angular estimations for all tested SNRs. The improvements in terms of SSL accuracies range from 23\% at -1 dB to 75\% at 20~dB when compared to MUSIC, and from 49\% at -1dB to 21\% at 20~dB when compared to SRP-PHAT. It is also interesting to note that for low-noise measurements (SNR~$\geq$~15 dB) in this reverberating environment, the SSL performances offered by MUSIC are degraded, since this subspace-based algorithm relies on the assumption that signal paths are uncorrelated or can be fully decorrelated, which does not hold true with early reflections in room acoustics, leaving DoA estimation a difficult problem in these situations for the wideband MUSIC method.

\subsection{Robustness to mismatch between the propagating environments in the learning and testing phases}

One of the main common pitfalls when using machine learning approaches resides in overfitting training data\cite{roelofs2019meta}. In order to avoid this phenomenon, the authors in\cite{perotin2019crnn} proposed the use of a training dataset composed by a large number of randomly sized rooms, which is an excellent way of circumventing these potential problems.\\

Still, it is interesting to study the robustness of the BeamLearning DNN, when tested in a totally different propagating environment than the one used for training. Using the same frozen network as in the previous subsection, trained using $\mathcal{D}_{\text{train}}^{\text{ Room1}}$, the frozen BeamLearning DNN is tested using 3600 sources positions in $\mathcal{D}_{\text{test speech}}^{\text{Room2}}$ in a $5 \times 5 \times 4$~m room, and the results are compared to those obtained using MUSIC and SRP-PHAT (see Tab.~\ref{tab:Prec_loca_salle}), for different SNR situations.\\

\begin{table}[ht]
	\caption{Mean absolute angular mismatch obtained for 3600 sources positions in $\mathcal{D}_{\text{test speech}}^{\text{Room2}}$, for different SNRs. The DNN has been pre-trained using  $\mathcal{D}_{\text{train}}^{\text{ Room1}}$, with random SNR~$\geq$5~dB. The best results are in bold.}
		\begin{tabularx}{\linewidth}{|b||s|s|s|s|s|s|}
						    \hline\hline
						    
			SNR (dB) & -1 & 2 & 5 & 10 & 15 & 20 \\ 
			\hline 
			SRP-PHAT & $53.7^{\circ}$ & $42.0^{\circ}$ & $30.5^{\circ}$  & $15.5^{\circ}$ & $9.2^{\circ}$ & $\bm{6.6}^{\circ}$ \\ 
			\hline 
			MUSIC & $\bm{41.9}^{\circ}$ & $36.7^{\circ}$ & $30.8^{\circ}$ & $26.9^{\circ}$ & $24.6^{\circ}$ & $25.0^{\circ}$ \\ 
			\hline 
			BeamLearning & $\bm{35.7}^{\circ}$ & $\bm{25.9}^{\circ}$ & $\bm{18.7^{\circ}}$ & $\bm{11.1}^{\circ}$ & $\bm{8.0}^{\circ}$ & $\bm{6.8}^{\circ}$ \\ 
									    \hline\hline
		\end{tabularx} 
	\label{tab:Prec_loca_salle} 
\end{table}

 This testing room exhibits a volume reduced by half, when compared to the one used during the training process. However, with a same reverberation time ($T_r = 0.5$~s), the testing room walls are therefore much more reflective, and the positions of the sources are much closer to the room boundaries than during the training process, which is commonly recognized as a difficult task for DoA estimation\cite{shen2020voice}. The degraded performances obtained using MUSIC in this testing room illustrates very well this observation. On the other hand, despite this significant change in the properties of the room between the training dataset and the testing dataset, the BeamLearning approach still offers significantly enhanced performances when compared to MUSIC and SRP-PHAT, except for a high SNR of 20~dB, where SRP-PHAT and BeamLearning allow equivalent SSL accuracies.

\subsection{Robustness to change of signal characteristics}

Similarly, we also investigate the influence of a mismatch between test signals emitted by the sources and the signals used during the training procedure. We use the same frozen network as in the previous subsection. As detailed in \ref{subsec:Datasets}, the signals emitted by the labelled sources in $\mathcal{D}_{\text{train}}^{\text{ Room1}}$ during the training process are composed of speech signals, symphonic music, and car horn honks. Even if the speech signals used for testing in the previous subsections were unseen during training, it is important to check if the performances remain satisfying when using a signal whose temporal and spectral dynamics are dissimilar to the ones used during the training process. This is the reason why Tab.~\ref{tab:Sig_rob} shows the localization performances obtained for SNRs of 5~dB and 15~dB using sound sources emitting dog bark excerpts from the UrbanSound8K dataset\cite{Salamon:UrbanSound:ACMMM:14}, along with the SSL results obtained with unseen voice signals.\\

The wideband MUSIC method allows a better localization of dog barks when compared to voice signals, mainly due to a smaller bandwidth occupied by the signal subspace when compared to the noise subspace. Still, the BeamLearning approach offers again better localization performances than both SRP-PHAT and wideband MUSIC in this situation. The analysis of these results allows to conclude that in each case, the BeamLearning approach outperforms MUSIC and SRP-PHAT in terms of SSL accuracy, and that the SSL performances observed with the frozen DNN do not degrade significantly when a signal with a different spectral and temporal characteristics is emitted by the sources to be located.

\begin{table}[ht]
		\caption{Mean absolute angular mismatch obtained using several unseen signals in an reverberant environment (RT~=~0.5~s), for different SNRs for 3600 sound sources positions. The DNN has been pre-trained using  $\mathcal{D}_{\text{train}}^{\text{ Room1}}$, with random SNR~$\geq$5~dB. The best results are in bold.}

		\begin{tabularx}{\linewidth}{|s||b|s|s|}
						    \hline\hline
			SNR & Localization approach & Voice signals & Dog bark \\ 
			\hline 
			& MUSIC &  $17.8^{\circ}$  &  $\bm{8.9}^{\circ}$ \\ 
			5 dB & SRP-PHAT & $17.4^{\circ}$ &   $16.9^{\circ}$ \\ 
			& BeamLearning & $\bm{8.2}^{\circ}$  & $\bm{6.7}^{\circ}$ \\ 
			\hline 
			& MUSIC &  $11.0^{\circ}$  &$6.4^{\circ}$\\ 
			15 dB & SRP-PHAT & $4.6^{\circ}$ &$6.1^{\circ}$\\  
			& BeamLearning  &  $\bm{3.4}^{\circ}$ &  $\bm{4.2}^{\circ}$ \\ 
								    \hline\hline	
		\end{tabularx} 
		
	\label{tab:Sig_rob} 
\end{table}

\subsection{Computational efficiency for realtime inference}

Since the proposed SSL method is a learning based method, the optimization of the $1.1\times10^6$ variables of the BeamLearning DNN represents a great amount of computational time (approx. 7 days on a Nvidia\textsuperscript{\small{\textregistered}} GTX 1080TI GPU card using the Tensorflow framework). This corresponds to a total of $10^6$ successive iterations for learning (approx. 434 epochs). Each of these learning iterations includes the calculation of the feed forward propagation, cross entropy loss, back-propagation, gradients computations, and variables updates using Adam, for each mini-batch. On this GPU architecture, and for each of the audio excerpts of 23.2 ms (1024 samples), the mean computation time for the learning process is therefore only 6 ms for the whole learning process involved. Given that some amount of these 6 ms are dedicated to the gradient optimizer operations, which are not needed for the inference with a frozen model, this gives confidence for a realtime inference.\\

\begin{table}[ht]
	\caption{Computation times for 3600 sources localizations}
		\begin{tabularx}{\linewidth}{|X|X|X|}
						    \hline\hline
			& Mean time (CPU)& Mean time (GPU)\\ 
			\hline 
			MUSIC & 45 min 40 s  & \textit{not implemented}  \\ 
			\hline 
			SRP-PHAT & 3 min 01 s  &  \textit{not implemented} \\ 
			\hline 
			BeamLearning & \textbf{2 min 47 s }  & \textbf{16.2 s} \\ 
									    \hline\hline
		\end{tabularx} 
	\label{tab:Calc_time} 
\end{table}

 In order to compare the computational time required by the three algorithms for an inference task, we benchmarked the SSL tasks detailed in section \ref{sec:Results} (see Table~\ref{tab:Calc_time}). The BeamLearning DNN has been tested both on a Nvidia\textsuperscript{\small{\textregistered}} GTX 1080TI GPU and a i7-6900K CPU (3.20GHz) with a mini-batch of size 1. On the other hand, MUSIC and SRP-PHAT have only been tested on the same CPU, since this is the only possible implementation proposed by\cite{pyroomacoustics}. In both cases, the BeamLearning approach allows to estimate the position of the sources in a faster way than the model approaches tested under the same conditions, with drastic improvements using a GPU computation.

\section{Conclusion}

In this paper, we have presented BeamLearning, a new deep learning approach for the localization of acoustic sources. The proposed DNN allows to achieve a 2D-DoA task in real-time from raw multichannel measurements on a microphone array. The proposed DNN architecture is partly inspired by the principle of filter and sum beamforming. In particular, the use of depthwise  separable atrous convolutions combined with pointwise convolutions allows to process the input audio data at multiple timescales. The BeamLearning network acts as a joint-feature learning process, that efficiently encodes the spatio-temporal information contained in raw measurements, and accomodates heavy measurement noise situations and reverberation. An extensive analysis of the BeamLearning approach using both a classification framework and a regression framework also allowed to show that the regression approach seems better suited when a fine angular resolution is seeked, with computation times that open up the possibility of precise and real-time localization using a deep learning approach, without any pre-processing of microphone array measurements. The analysis of the BeamLearning approach for SSL tasks in noisy and reverberating environments also proves that BeamLearning outperforms state of the art SRP-PHAT and MUSIC in almost each test situations.\\

In addition to the proposed network architecture, we also proposed the use of a fast and accurate RIR generator on GPU in order to build large realistic training datasets in an efficient way, using an image source method. Using these datasets, the proposed data augmentation process allowed to obtain increased robustness to measurement noise when compared to state of the art model-based SSL algorithms that were specifically crafted to handle this kind of degraded SNR situations. We also demonstrate that even if the propagation environment in which the localization is carried out significantly differs from the one used during the training phase, the BeamLearning approach remains quite robust.\\

The good SSL performances obtained for a 2D-DoA task motivates under development extensions of this work, which include realtime 3D-DoA and simultaneous localization / sound source recognition using the BeamLearning approach. Further developments may include the use of multi-room datasets and multi-source localization.

\section*{Acknowledgments}
This work has been partially supported through the Deeplomatics project funded by DGA/AID (ANR-18-ASTR-0008).







\begin{thebibliography}{10}
\def\enquote#1,{``#1,''}
\def\enxquote#1{``#1''}
\expandafter\ifx\csname url\endcsname\relax
  \def\url#1{\texttt{#1}}\fi
\expandafter\ifx\csname urlprefix\endcsname\relax\def\urlprefix{URL }\fi
\providecommand{\bibinfo}[2]{#2}
\def\plainquote#1{``#1''}
\providecommand{\noopsort}[1]{}
\providecommand{\switchargs}[2]{#2#1}
\providecommand{\dourl}[1]{\href{http://#1}{\nolinkurl{#1}}}
  \def\eatspace #1{#1}

\bibitem{steinberg1991neural}
\bibinfo{author}{B.~Z. Steinberg}, \bibinfo{author}{M.~J. Beran},
  \bibinfo{author}{S.~H. Chin}, and \bibinfo{author}{J.~H. Howard~Jr},
  \enquote{\bibinfo{title}{A neural network approach to source localization}},
  \bibinfo{journal}{The Journal of the Acoustical Society of America}
  \textbf{90}(4), \bibinfo{pages}{2081--2090} (\bibinfo{year}{1991}).

\bibitem{Guentchev1998learning}
\bibinfo{author}{K.~Guentchev} and \bibinfo{author}{J.~Weng},
  \enquote{\bibinfo{title}{Learning-based three dimensional sound localization
  using a compact non-coplanar array of microphones}}, in
  \emph{\bibinfo{booktitle}{Proc. AAAI Spring Symposium on Intelligent
  Environments}} (\bibinfo{year}{1998}).

\bibitem{weng2001three}
\bibinfo{author}{J.~Weng} and \bibinfo{author}{K.~Y. Guentchev},
  \enquote{\bibinfo{title}{Three-dimensional sound localization from a compact
  non-coplanar array of microphones using tree-based learning}},
  \bibinfo{journal}{The Journal of the Acoustical Society of America}
  \textbf{110}(1), \bibinfo{pages}{310--323} (\bibinfo{year}{2001}).

\bibitem{talmon2011supervised}
\bibinfo{author}{R.~Talmon}, \bibinfo{author}{I.~Cohen}, and
  \bibinfo{author}{S.~Gannot}, \enquote{\bibinfo{title}{Supervised source
  localization using diffusion kernels}}, in \emph{\bibinfo{booktitle}{2011
  IEEE workshop on applications of signal processing to audio and acoustics
  (WASPAA)}}, \bibinfo{organization}{IEEE} (\bibinfo{year}{2011}), pp.
  \bibinfo{pages}{245--248}.

\bibitem{deleforge2015acoustic}
\bibinfo{author}{A.~Deleforge}, \bibinfo{author}{F.~Forbes}, and
  \bibinfo{author}{R.~Horaud}, \enquote{\bibinfo{title}{Acoustic space learning
  for sound-source separation and localization on binaural manifolds}},
  \bibinfo{journal}{International journal of neural systems} \textbf{25}(1),
  \bibinfo{pages}{1440003} (\bibinfo{year}{2015}).

\bibitem{chakrabarty2019multi}
\bibinfo{author}{S.~Chakrabarty} and \bibinfo{author}{E.~A. Habets},
  \enquote{\bibinfo{title}{{Multi-speaker DOA estimation using deep
  convolutional networks trained with noise signals}}},  \bibinfo{journal}{IEEE
  Journal of Selected Topics in Signal Processing} \textbf{13}(1),
  \bibinfo{pages}{8--21} (\bibinfo{year}{2019}).

\bibitem{perotin2019crnn}
\bibinfo{author}{L.~Perotin}, \bibinfo{author}{R.~Serizel},
  \bibinfo{author}{E.~Vincent}, and \bibinfo{author}{A.~Guerin},
  \enquote{\bibinfo{title}{{CRNN-based multiple DoA estimation using acoustic
  intensity features for Ambisonics recordings}}},  \bibinfo{journal}{IEEE
  Journal of Selected Topics in Signal Processing} \textbf{13}(1),
  \bibinfo{pages}{22--33} (\bibinfo{year}{2019}).

\bibitem{adavanne2018sound}
\bibinfo{author}{S.~Adavanne}, \bibinfo{author}{A.~Politis},
  \bibinfo{author}{J.~Nikunen}, and \bibinfo{author}{T.~Virtanen},
  \enquote{\bibinfo{title}{{Sound Event Localization and Detection of
  Overlapping Sources Using Convolutional Recurrent Neural Networks}}},
  \bibinfo{journal}{IEEE Journ. of Selected Topics in Signal Processing}
  \textbf{13}(1), \bibinfo{pages}{34--48} (\bibinfo{year}{2019}).

\bibitem{brendel2019distributed}
\bibinfo{author}{A.~Brendel} and \bibinfo{author}{W.~Kellermann},
  \enquote{\bibinfo{title}{Distributed source localization in acoustic sensor
  networks using the coherent-to-diffuse power ratio}},  \bibinfo{journal}{IEEE
  Journal of Selected Topics in Signal Processing} \textbf{13}(1),
  \bibinfo{pages}{61--75} (\bibinfo{year}{2019}).

\bibitem{gannot2019introduction}
\bibinfo{author}{S.~Gannot}, \bibinfo{author}{M.~Haardt},
  \bibinfo{author}{W.~Kellermann}, and \bibinfo{author}{P.~Willett},
  \enquote{\bibinfo{title}{Introduction to the issue on acoustic source
  localization and tracking in dynamic real-life scenes}},
  \bibinfo{journal}{IEEE Journal of Selected Topics in Signal Processing}
  \textbf{13}(1), \bibinfo{pages}{3--7} (\bibinfo{year}{2019}).

\bibitem{lollmann2018locata}
\bibinfo{author}{H.~W. L{\"o}llmann}, \bibinfo{author}{C.~Evers},
  \bibinfo{author}{A.~Schmidt}, \bibinfo{author}{H.~Mellmann},
  \bibinfo{author}{H.~Barfuss}, \bibinfo{author}{P.~A. Naylor}, and
  \bibinfo{author}{W.~Kellermann}, \enquote{\bibinfo{title}{{The LOCATA
  challenge data corpus for acoustic source localization and tracking}}}, in
  \emph{\bibinfo{booktitle}{2018 IEEE 10th Sensor Array and Multichannel Signal
  Processing Workshop (SAM)}}, \bibinfo{organization}{IEEE}
  (\bibinfo{year}{2018}), pp. \bibinfo{pages}{410--414}.

\bibitem{pak2018locata}
\bibinfo{author}{J.~Pak} and \bibinfo{author}{J.~W. Shin},
  \enquote{\bibinfo{title}{{LOCATA Challenge: A Deep Neural Networks-Based
  Regression Approach for Direction-Of-Arrival Estimation}}}, in
  \emph{\bibinfo{booktitle}{Proc. of LOCATA Challenge Workshop-a satellite
  event of IWAENC}} (\bibinfo{year}{2018}).

\bibitem{liu2020source}
\bibinfo{author}{W.~Liu}, \bibinfo{author}{Y.~Yang}, \bibinfo{author}{M.~Xu},
  \bibinfo{author}{L.~L{\"u}}, \bibinfo{author}{Z.~Liu}, and
  \bibinfo{author}{Y.~Shi}, \enquote{\bibinfo{title}{Source localization in the
  deep ocean using a convolutional neural network}},  \bibinfo{journal}{The
  Journal of the Acoustical Society of America} \textbf{147}(4),
  \bibinfo{pages}{EL314--EL319} (\bibinfo{year}{2020}).

\bibitem{pak2019sound}
\bibinfo{author}{J.~Pak} and \bibinfo{author}{J.~W. Shin},
  \enquote{\bibinfo{title}{Sound localization based on phase difference
  enhancement using deep neural networks}},  \bibinfo{journal}{IEEE/ACM
  Transactions on Audio, Speech, and Language Processing} \textbf{27}(8),
  \bibinfo{pages}{1335--1345} (\bibinfo{year}{2019}).

\bibitem{comanducci2020source}
\bibinfo{author}{L.~Comanducci}, \bibinfo{author}{F.~Borra},
  \bibinfo{author}{P.~Bestagini}, \bibinfo{author}{F.~Antonacci},
  \bibinfo{author}{S.~Tubaro}, and \bibinfo{author}{A.~Sarti},
  \enquote{\bibinfo{title}{Source localization using distributed microphones in
  reverberant environments based on deep learning and ray space transform}},
  \bibinfo{journal}{IEEE/ACM Transactions on Audio, Speech, and Language
  Processing} \textbf{28}, \bibinfo{pages}{2238--2251} (\bibinfo{year}{2020}).

\bibitem{yasuda2020sound}
\bibinfo{author}{M.~Yasuda}, \bibinfo{author}{Y.~Koizumi},
  \bibinfo{author}{S.~Saito}, \bibinfo{author}{H.~Uematsu}, and
  \bibinfo{author}{K.~Imoto}, \enquote{\bibinfo{title}{{Sound Event
  Localization Based on Sound Intensity Vector Refined by Dnn-Based Denoising
  and Source Separation}}}, in \emph{\bibinfo{booktitle}{ICASSP 2020-2020 IEEE
  International Conference on Acoustics, Speech and Signal Processing
  (ICASSP)}}, \bibinfo{organization}{IEEE} (\bibinfo{year}{2020}), pp.
  \bibinfo{pages}{651--655}.

\bibitem{varzandeh2020exploiting}
\bibinfo{author}{R.~Varzandeh}, \bibinfo{author}{K.~Adilo{\u{g}}lu},
  \bibinfo{author}{S.~Doclo}, and \bibinfo{author}{V.~Hohmann},
  \enquote{\bibinfo{title}{{Exploiting Periodicity Features for Joint Detection
  and DOA Estimation of Speech Sources Using Convolutional Neural Networks}}},
  in \emph{\bibinfo{booktitle}{ICASSP 2020-2020 IEEE International Conference
  on Acoustics, Speech and Signal Processing (ICASSP)}},
  \bibinfo{organization}{IEEE} (\bibinfo{year}{2020}), pp.
  \bibinfo{pages}{566--570}.

\bibitem{gemba2019robust}
\bibinfo{author}{K.~L. Gemba}, \bibinfo{author}{S.~Nannuru}, and
  \bibinfo{author}{P.~Gerstoft}, \enquote{\bibinfo{title}{Robust ocean acoustic
  localization with sparse bayesian learning}},  \bibinfo{journal}{IEEE Journal
  of Selected Topics in Signal Processing} \textbf{13}(1),
  \bibinfo{pages}{49--60} (\bibinfo{year}{2019}).

\bibitem{liu2020multi}
\bibinfo{author}{Y.~Liu}, \bibinfo{author}{H.~Niu}, and
  \bibinfo{author}{Z.~Li}, \enquote{\bibinfo{title}{A multi-task learning
  convolutional neural network for source localization in deep ocean}},
  \bibinfo{journal}{The Journal of the Acoustical Society of America}
  \textbf{148}(2), \bibinfo{pages}{873--883} (\bibinfo{year}{2020}).

\bibitem{chakrabarty2017broadband}
\bibinfo{author}{S.~Chakrabarty} and \bibinfo{author}{E.~A. Habets},
  \enquote{\bibinfo{title}{{Broadband DOA estimation using convolutional neural
  networks trained with noise signals}}}, in
  \emph{\bibinfo{booktitle}{Applications of Signal Processing to Audio and
  Acoustics (WASPAA), 2017 IEEE Workshop on applications}}
  (\bibinfo{year}{2017}), pp. \bibinfo{pages}{136--140}.

\bibitem{Adavanne2019_DCASE}
\bibinfo{author}{S.~Adavanne}, \bibinfo{author}{A.~Politis}, and
  \bibinfo{author}{T.~Virtanen}, \enquote{\bibinfo{title}{{A Multi-room
  Reverberant Dataset for Sound Event Localization and Detection}}}, in
  \emph{\bibinfo{booktitle}{{Submitted to Detection and Classification of
  Acoustic Scenes and Events 2019 Workshop (DCASE2019)}}},
  \bibinfo{address}{Munich, Germany} (\bibinfo{year}{2019}),
  \url{https://arxiv.org/abs/1905.08546}.

\bibitem{hirvonen2015classification}
\bibinfo{author}{T.~Hirvonen}, \enquote{\bibinfo{title}{Classification of
  spatial audio location and content using convolutional neural networks}}, in
  \emph{\bibinfo{booktitle}{Audio Engineering Society Convention 138}},
  \bibinfo{organization}{Audio Engineering Society} (\bibinfo{year}{2015}).

\bibitem{he2018deep}
\bibinfo{author}{W.~He}, \bibinfo{author}{P.~Motlicek}, and
  \bibinfo{author}{J.-M. Odobez}, \enquote{\bibinfo{title}{Deep neural networks
  for multiple speaker detection and localization}}, in
  \emph{\bibinfo{booktitle}{2018 IEEE International Conference on Robotics and
  Automation (ICRA)}}, \bibinfo{organization}{IEEE} (\bibinfo{year}{2018}), pp.
  \bibinfo{pages}{74--79}.

\bibitem{sundar2020raw}
\bibinfo{author}{H.~Sundar}, \bibinfo{author}{W.~Wang},
  \bibinfo{author}{M.~Sun}, and \bibinfo{author}{C.~Wang},
  \enquote{\bibinfo{title}{Raw waveform based end-to-end deep convolutional
  network for spatial localization of multiple acoustic sources}}, in
  \emph{\bibinfo{booktitle}{ICASSP 2020-2020 IEEE International Conference on
  Acoustics, Speech and Signal Processing (ICASSP)}},
  \bibinfo{organization}{IEEE} (\bibinfo{year}{2020}), pp.
  \bibinfo{pages}{4642--4646}.

\bibitem{yalta2017sound}
\bibinfo{author}{N.~Yalta}, \bibinfo{author}{K.~Nakadai}, and
  \bibinfo{author}{T.~Ogata}, \enquote{\bibinfo{title}{Sound source
  localization using deep learning models}},  \bibinfo{journal}{Journal of
  Robotics and Mechatronics} \textbf{29}(1), \bibinfo{pages}{37--48}
  (\bibinfo{year}{2017}).

\bibitem{ma2017exploiting}
\bibinfo{author}{N.~Ma}, \bibinfo{author}{T.~May}, and \bibinfo{author}{G.~J.
  Brown}, \enquote{\bibinfo{title}{Exploiting deep neural networks and head
  movements for robust binaural localization of multiple sources in reverberant
  environments}},  \bibinfo{journal}{IEEE/ACM Transactions on Audio, Speech,
  and Language Processing} \textbf{25}(12), \bibinfo{pages}{2444--2453}
  (\bibinfo{year}{2017}).

\bibitem{nguyen2018autonomous}
\bibinfo{author}{Q.~Nguyen}, \bibinfo{author}{L.~Girin},
  \bibinfo{author}{G.~Bailly}, \bibinfo{author}{F.~Elisei}, and
  \bibinfo{author}{D.-C. Nguyen}, \enquote{\bibinfo{title}{Autonomous
  sensorimotor learning for sound source localization by a humanoid robot}}, in
  \emph{\bibinfo{booktitle}{Workshop on Crossmodal Learning for Intelligent
  Robotics in conjunction with IEEE/RSJ IROS}}.

\bibitem{sivasankaran2018keyword}
\bibinfo{author}{S.~Sivasankaran}, \bibinfo{author}{E.~Vincent}, and
  \bibinfo{author}{D.~Fohr}, \enquote{\bibinfo{title}{Keyword-based speaker
  localization: Localizing a target speaker in a multi-speaker environment}},
  in \emph{\bibinfo{booktitle}{Interspeech 2018 - 19th Annual Conference of the
  International Speech Communication Association}}.

\bibitem{vesperini2016neural}
\bibinfo{author}{F.~Vesperini}, \bibinfo{author}{P.~Vecchiotti},
  \bibinfo{author}{E.~Principi}, \bibinfo{author}{S.~Squartini}, and
  \bibinfo{author}{F.~Piazza}, \enquote{\bibinfo{title}{A neural network based
  algorithm for speaker localization in a multi-room environment}}, in
  \emph{\bibinfo{booktitle}{2016 IEEE 26th International Workshop on Machine
  Learning for Signal Processing (MLSP)}}, \bibinfo{organization}{IEEE}
  (\bibinfo{year}{2016}), pp. \bibinfo{pages}{1--6}.

\bibitem{ferguson2018sound}
\bibinfo{author}{E.~L. Ferguson}, \bibinfo{author}{S.~B. Williams}, and
  \bibinfo{author}{C.~T. Jin}, \enquote{\bibinfo{title}{Sound source
  localization in a multipath environment using convolutional neural
  networks}}, in \emph{\bibinfo{booktitle}{2018 IEEE International Conference
  on Acoustics, Speech and Signal Processing (ICASSP)}},
  \bibinfo{organization}{IEEE} (\bibinfo{year}{2018}), pp.
  \bibinfo{pages}{2386--2390}.

\bibitem{tang2019regression}
\bibinfo{author}{Z.~Tang}, \bibinfo{author}{J.~D. Kanu},
  \bibinfo{author}{K.~Hogan}, and \bibinfo{author}{D.~Manocha},
  \enquote{\bibinfo{title}{Regression and classification for
  direction-of-arrival estimation with convolutional recurrent neural
  networks}},  \bibinfo{journal}{arXiv preprint arXiv:1904.08452}
  (\bibinfo{year}{2019}).

\bibitem{salvati2018exploiting}
\bibinfo{author}{D.~Salvati}, \bibinfo{author}{C.~Drioli}, and
  \bibinfo{author}{G.~L. Foresti}, \enquote{\bibinfo{title}{Exploiting cnns for
  improving acoustic source localization in noisy and reverberant conditions}},
   \bibinfo{journal}{IEEE Transactions on Emerging Topics in Computational
  Intelligence} \textbf{2}(2), \bibinfo{pages}{103--116}
  (\bibinfo{year}{2018}).

\bibitem{takeda2016sound}
\bibinfo{author}{R.~Takeda} and \bibinfo{author}{K.~Komatani},
  \enquote{\bibinfo{title}{Sound source localization based on deep neural
  networks with directional activate function exploiting phase information}},
  in \emph{\bibinfo{booktitle}{Acoustics, Speech and Signal Processing
  (ICASSP), 2016 IEEE International Conference on Acoustics, speech and Signal
  Processing (ICASSP)}}, \bibinfo{organization}{IEEE} (\bibinfo{year}{2016}),
  pp. \bibinfo{pages}{405--409}.

\bibitem{yiwere2017distance}
\bibinfo{author}{M.~Yiwere} and \bibinfo{author}{E.~J. Rhee},
  \enquote{\bibinfo{title}{Distance estimation and localization of sound
  sources in reverberant conditions using deep neural networks}},
  \bibinfo{journal}{International Journal of Applied Engineering Research}
  \textbf{12}(22), \bibinfo{pages}{12384--12389} (\bibinfo{year}{2017}).

\bibitem{xiao2015learning}
\bibinfo{author}{X.~Xiao}, \bibinfo{author}{S.~Zhao},
  \bibinfo{author}{X.~Zhong}, \bibinfo{author}{D.~L. Jones},
  \bibinfo{author}{E.~S. Chng}, and \bibinfo{author}{H.~Li},
  \enquote{\bibinfo{title}{A learning-based approach to direction of arrival
  estimation in noisy and reverberant environments}}, in
  \emph{\bibinfo{booktitle}{2015 IEEE International Conference on Acoustics,
  Speech and Signal Processing (ICASSP)}} (\bibinfo{year}{2015}), pp.
  \bibinfo{pages}{2814--2818}.

\bibitem{suvorov2018deep}
\bibinfo{author}{D.~Suvorov}, \bibinfo{author}{G.~Dong}, and
  \bibinfo{author}{R.~Zhukov}, \enquote{\bibinfo{title}{Deep residual network
  for sound source localization in the time domain}},  \bibinfo{journal}{arXiv
  preprint arXiv:1808.06429}  (\bibinfo{year}{2018}).

\bibitem{vera2018towards}
\bibinfo{author}{J.~Vera-Diaz}, \bibinfo{author}{D.~Pizarro}, and
  \bibinfo{author}{J.~Macias-Guarasa}, \enquote{\bibinfo{title}{{Towards
  End-to-End Acoustic Localization Using Deep Learning: From Audio Signals to
  Source Position Coordinates}}},  \bibinfo{journal}{Sensors} \textbf{18}(10),
  \bibinfo{pages}{3418} (\bibinfo{year}{2018}).

\bibitem{huang2020time}
\bibinfo{author}{Y.~Huang}, \bibinfo{author}{X.~Wu}, and
  \bibinfo{author}{T.~Qu}, \enquote{\bibinfo{title}{A time-domain unsupervised
  learning based sound source localization method}}, in
  \emph{\bibinfo{booktitle}{2020 IEEE 3rd International Conference on
  Information Communication and Signal Processing (ICICSP)}},
  \bibinfo{organization}{IEEE} (\bibinfo{year}{2020}), pp.
  \bibinfo{pages}{26--32}.

\bibitem{comminiello2019quaternion}
\bibinfo{author}{D.~Comminiello}, \bibinfo{author}{M.~Lella},
  \bibinfo{author}{S.~Scardapane}, and \bibinfo{author}{A.~Uncini},
  \enquote{\bibinfo{title}{Quaternion convolutional neural networks for
  detection and localization of 3d sound events}}, in
  \emph{\bibinfo{booktitle}{2019 IEEE International Conference on Acoustics,
  Speech and Signal Processing (ICASSP)}}, \bibinfo{organization}{IEEE}
  (\bibinfo{year}{2019}), pp. \bibinfo{pages}{8533--8537}.

\bibitem{hu2020semi}
\bibinfo{author}{Y.~Hu}, \bibinfo{author}{P.~Samarasinghe},
  \bibinfo{author}{S.~Gannot}, and \bibinfo{author}{T.~Abhayapala},
  \enquote{\bibinfo{title}{Semi-supervised multiple source localization using
  relative harmonic coefficients under noisy and reverberant environments}},
  \bibinfo{journal}{IEEE/ACM Transactions on Audio, Speech, and Language
  Processing} \textbf{28}, \bibinfo{pages}{3108--3123} (\bibinfo{year}{2020}).

\bibitem{hu2020unsupervised}
\bibinfo{author}{Y.~Hu}, \bibinfo{author}{P.~N. Samarasinghe},
  \bibinfo{author}{T.~D. Abhayapala}, and \bibinfo{author}{S.~Gannot},
  \enquote{\bibinfo{title}{Unsupervised multiple source localization using
  relative harmonic coefficients}}, in \emph{\bibinfo{booktitle}{ICASSP
  2020-2020 IEEE International Conference on Acoustics, Speech and Signal
  Processing (ICASSP)}}, \bibinfo{organization}{IEEE} (\bibinfo{year}{2020}),
  pp. \bibinfo{pages}{571--575}.

\bibitem{stoter2018countnet}
\bibinfo{author}{F.-R. St{\"o}ter}, \bibinfo{author}{S.~Chakrabarty},
  \bibinfo{author}{B.~Edler}, and \bibinfo{author}{E.~A. Habets},
  \enquote{\bibinfo{title}{Countnet: Estimating the number of concurrent
  speakers using supervised learning}},  \bibinfo{journal}{IEEE/ACM
  Transactions on Audio, Speech, and Language Processing} \textbf{27}(2),
  \bibinfo{pages}{268--282} (\bibinfo{year}{2018}).

\bibitem{GrumiauxKGG20}
\bibinfo{author}{P.~Grumiaux}, \bibinfo{author}{S.~Kitic},
  \bibinfo{author}{L.~Girin}, and \bibinfo{author}{A.~Gu{\'{e}}rin},
  \enquote{\bibinfo{title}{High-resolution speaker counting in reverberant
  rooms using {CRNN} with ambisonics features}}, in
  \emph{\bibinfo{booktitle}{28th European Signal Processing Conference,
  {EUSIPCO} 2020, Amsterdam, Netherlands, January 18-21, 2021}},
  \bibinfo{publisher}{{IEEE}} (\bibinfo{year}{2020}), pp.
  \bibinfo{pages}{71--75}.

\bibitem{bianco2019machine}
\bibinfo{author}{M.~J. Bianco}, \bibinfo{author}{P.~Gerstoft},
  \bibinfo{author}{J.~Traer}, \bibinfo{author}{E.~Ozanich},
  \bibinfo{author}{M.~A. Roch}, \bibinfo{author}{S.~Gannot}, and
  \bibinfo{author}{C.-A. Deledalle}, \enquote{\bibinfo{title}{Machine learning
  in acoustics: Theory and applications}},  \bibinfo{journal}{The Journal of
  the Acoustical Society of America} \textbf{146}(5),
  \bibinfo{pages}{3590--3628} (\bibinfo{year}{2019}).

\bibitem{dieleman2014end}
\bibinfo{author}{S.~Dieleman} and \bibinfo{author}{B.~Schrauwen},
  \enquote{\bibinfo{title}{End-to-end learning for music audio}}, in
  \emph{\bibinfo{booktitle}{2014 IEEE International Conference on Acoustics,
  Speech and Signal Processing (ICASSP)}}, \bibinfo{organization}{IEEE}
  (\bibinfo{year}{2014}), pp. \bibinfo{pages}{6964--6968}.

\bibitem{dai2017very}
\bibinfo{author}{W.~Dai}, \bibinfo{author}{C.~Dai}, \bibinfo{author}{S.~Qu},
  \bibinfo{author}{J.~Li}, and \bibinfo{author}{S.~Das},
  \enquote{\bibinfo{title}{Very deep convolutional neural networks for raw
  waveforms}}, in \emph{\bibinfo{booktitle}{2017 IEEE International Conference
  on Acoustics, Speech and Signal Processing (ICASSP)}},
  \bibinfo{organization}{IEEE} (\bibinfo{year}{2017}), pp.
  \bibinfo{pages}{421--425}.

\bibitem{sainath2015learning}
\bibinfo{author}{T.~N. Sainath}, \bibinfo{author}{R.~J. Weiss},
  \bibinfo{author}{A.~Senior}, \bibinfo{author}{K.~W. Wilson}, and
  \bibinfo{author}{O.~Vinyals}, \enquote{\bibinfo{title}{{Learning the speech
  front-end with raw waveform CLDNNs}}}, in \emph{\bibinfo{booktitle}{Sixteenth
  Annual Conference of the International Speech Communication Association}}
  (\bibinfo{year}{2015}).

\bibitem{lee2018samplecnn}
\bibinfo{author}{J.~Lee}, \bibinfo{author}{J.~Park}, \bibinfo{author}{K.~L.
  Kim}, and \bibinfo{author}{J.~Nam}, \enquote{\bibinfo{title}{Samplecnn:
  End-to-end deep convolutional neural networks using very small filters for
  music classification}},  \bibinfo{journal}{Applied Sciences} \textbf{8}(1),
  \bibinfo{pages}{150} (\bibinfo{year}{2018}).

\bibitem{bavu2019timescalenet}
\bibinfo{author}{{\'E}.~Bavu}, \bibinfo{author}{A.~Ramamonjy},
  \bibinfo{author}{H.~Pujol}, and \bibinfo{author}{A.~Garcia},
  \enquote{\bibinfo{title}{{TimeScaleNet : a Multiresolution Approach for Raw
  Audio Recognition using Learnable Biquadratic IIR Filters and Residual
  Networks of Depthwise-Separable One-Dimensional Atrous Convolutions}}},
  \bibinfo{journal}{IEEE Journ. of Selected Topics in Signal Processing}
  \textbf{13}(2), \bibinfo{pages}{220--235} (\bibinfo{year}{2019}).
  
  \bibitem{chollet2017xception}
  \bibinfo{author}{F.~Chollet}, \enquote{\bibinfo{title}{Deep learning with depthwise separable convolutions}}, in
  \emph{\bibinfo{booktitle}{Proceedings of the IEEE conference on computer vision and pattern recognition}}, \bibinfo{organization}{IEEE}
  (\bibinfo{year}{2017}), pp. \bibinfo{pages}{1251--1258}.
  

\bibitem{ravanelli2018speaker}
\bibinfo{author}{M.~Ravanelli} and \bibinfo{author}{Y.~Bengio},
  \enquote{\bibinfo{title}{Speaker recognition from raw waveform with
  sincnet}}, in \emph{\bibinfo{booktitle}{2018 IEEE Spoken Language Technology
  Workshop (SLT)}}, \bibinfo{organization}{IEEE} (\bibinfo{year}{2018}), pp.
  \bibinfo{pages}{1021--1028}.

\bibitem{oord2016wavenet}
\bibinfo{author}{A.~van~den Oord}, \bibinfo{author}{S.~Dieleman},
  \bibinfo{author}{H.~Zen}, \bibinfo{author}{K.~Simonyan},
  \bibinfo{author}{O.~Vinyals}, \bibinfo{author}{A.~Graves},
  \bibinfo{author}{N.~Kalchbrenner}, \bibinfo{author}{A.~Senior}, and
  \bibinfo{author}{K.~Kavukcuoglu}, \enquote{\bibinfo{title}{Wavenet: A
  generative model for raw audio}},  \bibinfo{journal}{arXiv preprint
  arXiv:1609.03499}  (\bibinfo{year}{2016}).

\bibitem{kaiser2017depthwise}
\bibinfo{author}{L.~Kaiser}, \bibinfo{author}{A.~N. Gomez}, and
  \bibinfo{author}{F.~Chollet}, \enquote{\bibinfo{title}{Depthwise separable
  convolutions for neural machine translation}},  \bibinfo{journal}{arXiv
  preprint arXiv:1706.03059}  (\bibinfo{year}{2017}).

\bibitem{rethage2018wavenet}
\bibinfo{author}{D.~Rethage}, \bibinfo{author}{J.~Pons}, and
  \bibinfo{author}{X.~Serra}, \enquote{\bibinfo{title}{A wavenet for speech
  denoising}}, in \emph{\bibinfo{booktitle}{2018 IEEE International Conference
  on Acoustics, Speech and Signal Processing (ICASSP)}},
  \bibinfo{organization}{IEEE} (\bibinfo{year}{2018}), pp.
  \bibinfo{pages}{5069--5073}.

\bibitem{perotin2019regression}
\bibinfo{author}{L.~Perotin}, \bibinfo{author}{A.~D{\'e}fossez},
  \bibinfo{author}{E.~Vincent}, \bibinfo{author}{R.~Serizel}, and
  \bibinfo{author}{A.~Gu{\'e}rin}, \enquote{\bibinfo{title}{Regression versus
  classification for neural network based audio source localization}}, in
  \emph{\bibinfo{booktitle}{2019 IEEE Workshop on Applications of Signal
  Processing to Audio and Acoustics (WASPAA)}}, \bibinfo{organization}{IEEE}
  (\bibinfo{year}{2019}), pp. \bibinfo{pages}{343--347}.

\bibitem{he2019adaptation}
\bibinfo{author}{W.~{He}}, \bibinfo{author}{P.~{Motlicek}}, and
  \bibinfo{author}{J.~{Odobez}}, \enquote{\bibinfo{title}{Adaptation of
  multiple sound source localization neural networks with weak supervision and
  domain-adversarial training}}, in \emph{\bibinfo{booktitle}{ICASSP 2019 -
  2019 IEEE International Conference on Acoustics, Speech and Signal Processing
  (ICASSP)}} (\bibinfo{year}{2019}), pp. \bibinfo{pages}{770--774}.

\bibitem{lecomte2016fifty}
\bibinfo{author}{P.~Lecomte}, \bibinfo{author}{P.-A. Gauthier},
  \bibinfo{author}{C.~Langrenne}, \bibinfo{author}{A.~Berry}, and
  \bibinfo{author}{A.~Garcia}, \enquote{\bibinfo{title}{A fifty-node lebedev
  grid and its applications to ambisonics}},  \bibinfo{journal}{Journal of the
  Audio Engineering Society} \textbf{64}(11), \bibinfo{pages}{868--881}
  (\bibinfo{year}{2016}).

\bibitem{lecomteambitools}
\bibinfo{author}{P.~Lecomte}, \enquote{\bibinfo{title}{Ambitools: Tools for
  sound field synthesis with higher order ambisonics-v1. 0}}, in
  \emph{\bibinfo{booktitle}{1st international Faust Conference (IFC-18)}}.

\bibitem{pujol2019source}
\bibinfo{author}{H.~Pujol}, \bibinfo{author}{E.~Bavu}, and
  \bibinfo{author}{A.~Garcia}, \enquote{\bibinfo{title}{{Source localization in
  reverberant rooms using Deep Learning and microphone arrays}}}, in
  \emph{\bibinfo{booktitle}{{23rd International Congress on Acoustics (ICA 2019
  Aachen)}}}, \bibinfo{address}{Aachen, Germany} (\bibinfo{year}{2019}), pp.
  \bibinfo{pages}{6929--6936}.

\bibitem{allen1979image}
\bibinfo{author}{J.~B. Allen} and \bibinfo{author}{D.~A. Berkley},
  \enquote{\bibinfo{title}{Image method for efficiently simulating small-room
  acoustics}},  \bibinfo{journal}{The Journal of the Acoustical Society of
  America} \textbf{65}(4), \bibinfo{pages}{943--950} (\bibinfo{year}{1979}).

\bibitem{pyroomacoustics}
\bibinfo{author}{R.~Scheibler}, \bibinfo{author}{E.~Bezzam}, and
  \bibinfo{author}{I.~Dokmani{\'c}}, \enquote{\bibinfo{title}{Pyroomacoustics:
  A python package for audio room simulation and array processing algorithms}},
  in \emph{\bibinfo{booktitle}{2018 IEEE International Conference on Acoustics,
  Speech and Signal Processing (ICASSP)}}, \bibinfo{organization}{IEEE}
  (\bibinfo{year}{2018}), pp. \bibinfo{pages}{351--355}.

\bibitem{tensorflow2015-whitepaper}
\bibinfo{author}{M.~Abadi}, \bibinfo{author}{A.~Agarwal}, \emph{et~al.},
  \enxquote{\bibinfo{title}{{TensorFlow}: Large-scale machine learning on
  heterogeneous systems}}  (\bibinfo{year}{2015}),
  \url{http://tensorflow.org/} \bibinfo{note}{software available from
  tensorflow.org}.

\bibitem{smithdigitalfilters}
\bibinfo{author}{J.~O. Smith}, \emph{\bibinfo{title}{Introduction to Digital
  Filters with Audio Applications}}  (\bibinfo{publisher}{W3K Publishing},
  \bibinfo{year}{2007}).

\bibitem{brandstein2013microphone}
\bibinfo{author}{M.~Brandstein} and \bibinfo{author}{D.~Ward},
  \emph{\bibinfo{title}{Microphone arrays: signal processing techniques and
  applications}}  (\bibinfo{publisher}{Springer-Verlag},
  \bibinfo{address}{Berlin, Germany}, \bibinfo{year}{2001}).
  
    \bibitem{holschneider1990real}
\bibinfo{author}{M.~Holschneider}, \bibinfo{author}{R.~Kronland-Martinet},
  \bibinfo{author}{J.~Morlet}, and \bibinfo{author}{P.~Tchamitchian},
  \enquote{\bibinfo{title}{A real-time algorithm for signal analysis with the
  help of the wavelet transform}}, in \emph{\bibinfo{booktitle}{Wavelets}}
  (\bibinfo{publisher}{Springer}, \bibinfo{year}{1990}), pp.
  \bibinfo{pages}{286--297}.

\bibitem{he2016identity}
\bibinfo{author}{K.~He}, \bibinfo{author}{X.~Zhang}, \bibinfo{author}{S.~Ren},
  and \bibinfo{author}{J.~Sun}, \enquote{\bibinfo{title}{Identity mappings in
  deep residual networks}}, in \emph{\bibinfo{booktitle}{European conference on
  computer vision ECCV'16}}, \bibinfo{organization}{Springer}
  (\bibinfo{year}{2016}), Vol. \bibinfo{volume}{9908}, pp.
  \bibinfo{pages}{630--645}.

\bibitem{he2016deep}
\bibinfo{author}{K.~He}, \bibinfo{author}{X.~Zhang}, \bibinfo{author}{S.~Ren},
  and \bibinfo{author}{J.~Sun}, \enquote{\bibinfo{title}{Deep residual learning
  for image recognition}}, in \emph{\bibinfo{booktitle}{Proceedings of the IEEE
  conference on computer vision and pattern recognition (CVPR)}}
  (\bibinfo{year}{2016}), pp. \bibinfo{pages}{770--778}.

\bibitem{srivastava2015training}
\bibinfo{author}{R.~K. Srivastava}, \bibinfo{author}{K.~Greff}, and
  \bibinfo{author}{J.~Schmidhuber}, \enquote{\bibinfo{title}{Training very deep
  networks}}, in \emph{\bibinfo{booktitle}{NIPS'15: Proceedings of the 28th
  Internatinal Conference on Neuronal information processing systems}}
  (\bibinfo{year}{2015}), pp. \bibinfo{pages}{2377--2385}.

\bibitem{ba2016layer}
\bibinfo{author}{J.~L. Ba}, \bibinfo{author}{J.~R. Kiros}, and
  \bibinfo{author}{G.~E. Hinton}, \enquote{\bibinfo{title}{Layer
  normalization}},  \bibinfo{journal}{arXiv preprint arXiv:1607.06450}
  (\bibinfo{year}{2016}).

\bibitem{ioffe2015batch}
\bibinfo{author}{S.~Ioffe} and \bibinfo{author}{C.~Szegedy},
  \enquote{\bibinfo{title}{Batch normalization: Accelerating deep network
  training by reducing internal covariate shift}},  \bibinfo{journal}{arXiv
  preprint arXiv:1502.03167}  (\bibinfo{year}{2015}).

\bibitem{ioffe2017batch}
\bibinfo{author}{S.~Ioffe}, \enquote{\bibinfo{title}{Batch renormalization:
  Towards reducing minibatch dependence in batch-normalized models}}, in
  \emph{\bibinfo{booktitle}{Advances in neural information processing systems}}
  (\bibinfo{year}{2017}), pp. \bibinfo{pages}{1945--1953}.

\bibitem{klambauer2017self}
\bibinfo{author}{G.~Klambauer}, \bibinfo{author}{T.~Unterthiner},
  \bibinfo{author}{A.~Mayr}, and \bibinfo{author}{S.~Hochreiter},
  \enquote{\bibinfo{title}{Self-normalizing neural networks}}, in
  \emph{\bibinfo{booktitle}{Advances in Neural Information Processing Systems
  30}} (\bibinfo{year}{2017}), pp. \bibinfo{pages}{972--981}.

\bibitem{kingma2014adam}
\bibinfo{author}{D.~P. Kingma} and \bibinfo{author}{J.~Ba},
  \enquote{\bibinfo{title}{Adam: A method for stochastic optimization}},
  \bibinfo{journal}{arXiv preprint arXiv:1412.6980}  (\bibinfo{year}{2014}).

\bibitem{lehmann2007reverberation}
\bibinfo{author}{E.~A. Lehmann}, \bibinfo{author}{A.~M. Johansson}, and
  \bibinfo{author}{S.~Nordholm}, \enquote{\bibinfo{title}{Reverberation-time
  prediction method for room impulse responses simulated with the image-source
  model}}, in \emph{\bibinfo{booktitle}{2007 IEEE Workshop on Applications of
  Signal Processing to Audio and Acoustics}}, \bibinfo{organization}{IEEE}
  (\bibinfo{year}{2007}), pp. \bibinfo{pages}{159--162}.

\bibitem{lehmann2008prediction}
\bibinfo{author}{E.~A. Lehmann} and \bibinfo{author}{A.~M. Johansson},
  \enquote{\bibinfo{title}{Prediction of energy decay in room impulse responses
  simulated with an image-source model}},  \bibinfo{journal}{The Journal of the
  Acoustical Society of America} \textbf{124}(1), \bibinfo{pages}{269--277}
  (\bibinfo{year}{2008}).

\bibitem{lokki2008recording}
\bibinfo{author}{T.~Lokki}, \bibinfo{author}{J.~Patynen}, and
  \bibinfo{author}{V.~Pulkki}, \enquote{\bibinfo{title}{Recording of anechoic
  symphony music}},  \bibinfo{journal}{Journal of the Acoustical Society of
  America} \textbf{123}(5), \bibinfo{pages}{3936--3936} (\bibinfo{year}{2008}).

\bibitem{Salamon:UrbanSound:ACMMM:14}
\bibinfo{author}{J.~Salamon}, \bibinfo{author}{C.~Jacoby}, and
  \bibinfo{author}{J.~P. Bello}, \enquote{\bibinfo{title}{A dataset and
  taxonomy for urban sound research}}, in \emph{\bibinfo{booktitle}{22nd {ACM}
  International Conference on Multimedia (ACM-MM'14)}},
  \bibinfo{address}{Orlando, FL, USA}, pp. \bibinfo{pages}{1041--1044}.

\bibitem{schmidt1986multiple}
\bibinfo{author}{R.~Schmidt}, \enquote{\bibinfo{title}{Multiple emitter
  location and signal parameter estimation}},  \bibinfo{journal}{IEEE
  transactions on antennas and propagation} \textbf{34}(3),
  \bibinfo{pages}{276--280} (\bibinfo{year}{1986}).

\bibitem{dibiase2001robust}
\bibinfo{author}{J.~H. DiBiase}, \bibinfo{author}{H.~F. Silverman}, and
  \bibinfo{author}{M.~S. Brandstein}, \enquote{\bibinfo{title}{Robust
  localization in reverberant rooms}}, in \emph{\bibinfo{booktitle}{Microphone
  Arrays}}, Chap.~\bibinfo{chapter}{8}, pp. \bibinfo{pages}{157--180}.

\bibitem{sokolova2009systematic}
\bibinfo{author}{M.~Sokolova} and \bibinfo{author}{G.~Lapalme},
  \enquote{\bibinfo{title}{A systematic analysis of performance measures for
  classification tasks}},  \bibinfo{journal}{Information processing \&
  management} \textbf{45}(4), \bibinfo{pages}{427--437} (\bibinfo{year}{2009}).

\bibitem{do2010srp}
\bibinfo{author}{H.~Do} and \bibinfo{author}{H.~F. Silverman},
  \enquote{\bibinfo{title}{{SRP-PHAT methods of locating simultaneous multiple
  talkers using a frame of microphone array data}}}, in
  \emph{\bibinfo{booktitle}{2010 IEEE International Conference on Acoustics,
  Speech and Signal Processing}}, \bibinfo{organization}{IEEE}
  (\bibinfo{year}{2010}), pp. \bibinfo{pages}{125--128}.

\bibitem{badali2009evaluating}
\bibinfo{author}{A.~Badali}, \bibinfo{author}{J.-M. Valin},
  \bibinfo{author}{F.~Michaud}, and \bibinfo{author}{P.~Aarabi},
  \enquote{\bibinfo{title}{Evaluating real-time audio localization algorithms
  for artificial audition in robotics}}, in \emph{\bibinfo{booktitle}{2009
  IEEE/RSJ International Conference on Intelligent Robots and Systems}},
  \bibinfo{organization}{IEEE} (\bibinfo{year}{2009}), pp.
  \bibinfo{pages}{2033--2038}.

\bibitem{cobos2010modified}
\bibinfo{author}{M.~Cobos}, \bibinfo{author}{A.~Marti}, and
  \bibinfo{author}{J.~J. Lopez}, \enquote{\bibinfo{title}{{A modified SRP-PHAT
  functional for robust real-time sound source localization with scalable
  spatial sampling}}},  \bibinfo{journal}{IEEE Signal Processing Letters}
  \textbf{18}(1), \bibinfo{pages}{71--74} (\bibinfo{year}{2010}).

\bibitem{dmochowski2010steered}
\bibinfo{author}{J.~P. Dmochowski} and \bibinfo{author}{J.~Benesty},
  \enquote{\bibinfo{title}{Steered beamforming approaches for acoustic source
  localization}}, in \emph{\bibinfo{booktitle}{Speech processing in modern
  communication}}  (\bibinfo{publisher}{Springer}, \bibinfo{year}{2010}), pp.
  \bibinfo{pages}{307--337}.

\bibitem{zhang2008does}
\bibinfo{author}{C.~Zhang}, \bibinfo{author}{D.~Flor{\^e}ncio}, and
  \bibinfo{author}{Z.~Zhang}, \enquote{\bibinfo{title}{Why does {PHAT} work
  well in lownoise, reverberative environments?}}, in
  \emph{\bibinfo{booktitle}{2008 IEEE International Conference on Acoustics,
  Speech and Signal Processing}}, \bibinfo{organization}{IEEE}
  (\bibinfo{year}{2008}), pp. \bibinfo{pages}{2565--2568}.

\bibitem{argentieri2007broadband}
\bibinfo{author}{S.~Argentieri} and \bibinfo{author}{P.~Danes},
  \enquote{\bibinfo{title}{{Broadband variations of the MUSIC high-resolution
  method for sound source localization in robotics}}}, in
  \emph{\bibinfo{booktitle}{2007 IEEE/RSJ International Conference on
  Intelligent Robots and Systems}}, \bibinfo{organization}{IEEE}
  (\bibinfo{year}{2007}), pp. \bibinfo{pages}{2009--2014}.

\bibitem{dmochowski2007broadband}
\bibinfo{author}{J.~P. Dmochowski}, \bibinfo{author}{J.~Benesty}, and
  \bibinfo{author}{S.~Affes}, \enquote{\bibinfo{title}{{Broadband MUSIC:
  Opportunities and challenges for multiple source localization}}}, in
  \emph{\bibinfo{booktitle}{2007 IEEE Workshop on Applications of Signal
  Processing to Audio and Acoustics}}, \bibinfo{organization}{IEEE}
  (\bibinfo{year}{2007}), pp. \bibinfo{pages}{18--21}.

\bibitem{ishi2009evaluation}
\bibinfo{author}{C.~T. Ishi}, \bibinfo{author}{O.~Chatot},
  \bibinfo{author}{H.~Ishiguro}, and \bibinfo{author}{N.~Hagita},
  \enquote{\bibinfo{title}{{Evaluation of a MUSIC-based real-time sound
  localization of multiple sound sources in real noisy environments}}}, in
  \emph{\bibinfo{booktitle}{2009 IEEE/RSJ International Conference on
  Intelligent Robots and Systems}}, \bibinfo{organization}{IEEE}
  (\bibinfo{year}{2009}), pp. \bibinfo{pages}{2027--2032}.

\bibitem{zhang2009robust}
\bibinfo{author}{J.~X. Zhang}, \bibinfo{author}{M.~G. Christensen},
  \bibinfo{author}{J.~Dahl}, \bibinfo{author}{S.~H. Jensen}, and
  \bibinfo{author}{M.~Moonen}, \enquote{\bibinfo{title}{{Robust implementation
  of the MUSIC algorithm}}}, in \emph{\bibinfo{booktitle}{2009 IEEE
  International Conference on Acoustics, Speech and Signal Processing}},
  \bibinfo{organization}{IEEE} (\bibinfo{year}{2009}), pp.
  \bibinfo{pages}{3037--3040}.

\bibitem{roelofs2019meta}
\bibinfo{author}{R.~Roelofs}, \bibinfo{author}{V.~Shankar},
  \bibinfo{author}{B.~Recht}, \bibinfo{author}{S.~Fridovich-Keil},
  \bibinfo{author}{M.~Hardt}, \bibinfo{author}{J.~Miller}, and
  \bibinfo{author}{L.~Schmidt}, \enquote{\bibinfo{title}{A meta-analysis of
  overfitting in machine learning}},  \bibinfo{journal}{Advances in Neural
  Information Processing Systems} \textbf{32}, \bibinfo{pages}{9179--9189}
  (\bibinfo{year}{2019}).

\bibitem{shen2020voice}
\bibinfo{author}{S.~Shen}, \bibinfo{author}{D.~Chen}, \bibinfo{author}{Y.-L.
  Wei}, \bibinfo{author}{Z.~Yang}, and \bibinfo{author}{R.~R. Choudhury},
  \enquote{\bibinfo{title}{Voice localization using nearby wall reflections}},
  in \emph{\bibinfo{booktitle}{Proceedings of the 26th Annual International
  Conference on Mobile Computing and Networking}} (\bibinfo{year}{2020}), pp.
  \bibinfo{pages}{1--14}.
  


\end{thebibliography}

\typeout{}

\end{document}